\documentclass[nofootinbib,floatfix,twocolumn]{revtex4}
\usepackage{empheq} 
\usepackage{mathrsfs}
\usepackage{verbatim}
\usepackage{amssymb}
\usepackage{graphicx}
\usepackage{subfigure}
\usepackage{color}
\usepackage{rotating}

\usepackage[mathscr]{euscript}
\usepackage{amsmath}

\def\H{H}
\def\bhx{{\bf {\hat X}}}
\def\bhy{{\bf {\hat Y}}}
\def\bhPhi{{\bf {\hat \Phi}}}
\def\bhPsi{{\bf {\hat \Psi}}}
\def\bhi{{\bf {\hat i}}}
\def\bhjy{{\bf {\hat j}}}
\def\bhk{{\bf {\hat k}}}

\def\br{{\bf r}}

\def\bp{{\bf p}}
\def\bn{{\bf n}}
\def\bl{{\bf L}}
\def\bhl{{\bf {\hat L}}}
\def\bs{{\bf S}}
\def\bhs{{\bf {\hat S}}}
\def\bj{{\bf J}}
\def\bhj{{\bf {\hat J}}}
\def\bn{{\bf {\hat n}}}

\def\bhvarphi{{\bf {\hat {\mathbf \varphi}}}}

\def\be{\begin{equation}}
\def\ee{\end{equation}}
\def\ba{\begin{eqnarray}}
\def\ea{\end{eqnarray}}

\newcommand{\half}[0]{\frac{1}{2}}

\begin{document}
\bibliographystyle{apsrev}

\title{Dynamics of Black Hole Pairs I:\par Periodic Tables}

\author{Janna Levin${}^{*,!}$ and Becky Grossman${}^{**}$}
\affiliation{${}^{*}$Department of Physics and Astronomy, Barnard
College of Columbia University, 3009 Broadway, New York, NY 10027 }
\affiliation{${}^{!}$Institute for Strings, Cosmology and Astroparticle
  Physics, Columbia University, New York, NY 10027}
\affiliation{${}^{**}$Physics Department, Columbia University,
New York, NY 10027}
\affiliation{ janna@astro.columbia.edu}
\affiliation{ becky@phys.columbia.edu }

\widetext

\begin{abstract}
Although the orbits of comparable mass, spinning black holes seem to
defy simple decoding, we find a means to decipher all such orbits. The
dynamics is complicated by extreme perihelion precession compounded by
spin-induced precession.  We are able to quantitatively define and
describe the fully three dimensional motion of comparable mass
binaries with one black hole spinning and expose an underlying
simplicity. To do so, we untangle the dynamics by capturing the motion
in the orbital plane and explicitly separate out the precession of the
plane itself. Our system is defined by the 3PN Hamiltonian plus
spin-orbit coupling for one spinning black hole with a non-spinning
companion. Our results are twofold: (1) We derive highly simplified
equations of motion in a non-orthogonal orbital basis, and (2) we
define a complete taxonomy for fully three-dimensional orbits.  More
than just a naming system, the taxonomy provides unambiguous and
quantitative descriptions of the orbits, including a determination of
the zoom-whirliness of any given orbit. Through a correspondence with
the rationals, we are able to show that zoom-whirl behavior is
prevalent in comparable mass binaries in the strong-field regime, as
it is for extreme-mass-ratio binaries in the strong-field. A first
significant conclusion that can be drawn from this analysis is that
all generic orbits in the final stages of inspiral under gravitational
radiation losses are characterized by precessing clovers with few
leaves and that no orbit will behave like the tightly precessing
ellipse of Mercury. The gravitational waveform produced by these
low-leaf clovers will reflect the natural harmonics of the orbital
basis -- harmonics that, importantly, depend only on radius. The
significance for gravitational wave astronomy will depend on the
number of windings the pair executes in the strong-field regime and
could be more conspicuous for intermediate mass pairs than for stellar
mass pairs. The 3PN system studied provides an example of a general
method that can be applied to any effective description of black hole
pairs.

\end{abstract}

\maketitle

\vfill\eject

At first glance, the orbits of a black hole pair resist coherent description. Dynamically, black hole
pairs involve non-linear relativistic effects leading to an extreme
form of perihelion precession, coined zoom-whirl
behavior,
as well as spin precession that in turn drives orbits
out of a plane. The three-dimensional precessions fill out a tangled
path that
shapes the gravitational waves both LIGO \cite{ligo} and LISA
\cite{lisa} were designed to observe
\cite{{apostolatos1994},{kidder1995},{grandclement2003},{faye2006},{blanchet2006},{vecchio2004},{lang2006}}. 
Despite appearances, we show the path can be untangled and a coherent
description of fully three-dimensional precessing orbits proves
entirely possible for one spinning black hole, one non-spinning.

The spacetime around two orbiting black holes eludes analytic description. 
While the impressive breakthroughs of
numerical relativity
\cite{{pretorius2006},{herrmann2007},{campanelli2007},{campanelli2006},{baker2005},{marronetti2007},{scheel2006}}
promise to resolve the final plunge of a black
hole pair, computational expense relegates the majority of the
inspiral to analytic methods.
Several groups have gone
to great pains to build a Hamiltonian formulation for 
two black holes in a Post-Newtonian (PN) approximation for analytic
computation of orbits and the gravitational waves they generate. As a
contribution to this great campaign, 
we use the conservative 3PN Hamiltonian plus spin-orbit couplings to
deconstruct the full three-dimensional dynamics of black hole
pairs, in the absence of radiation reaction. Our binaries are composed
of one spinning black hole and one
non-spinning black hole of any mass ratio.

The complicated three-dimensional motion can be beautifully decomposed
into two-dimensional motion in an orbital plane with a precession of
the orbital plane superposed \cite{apostolatos1994}. Through this modular decomposition, we
are able to define a complete taxonomy of all three-dimensional orbits
in terms of orbits that are closed in the orbital plane. 
Further, we find the spectrum of orbits for a given black hole
system. Importantly, the spectrum in the strong-field regime
shows zoom-whirl behavior
during which an orbit
sweeps out to apastron and back
in a zoom followed 
by a multiplicity of nearly circular whirls around the
center of mass. It must be emphasized that our results prove
zoom-whirl behavior is ubiquitous even for comparable mass binaries
and unambiguously quantified in our taxonomy.
Importantly, zoom-whirl motion has already been observed in numerical
relativity \cite{pretorius2007}.

From the outset, we acknowledge that the PN approximation is pushed to
its breaking point in the strong-field regime where zoom-whirl
behavior is most prevalent. However, the method we advocate --
locating a periodic skeleton in an orbital basis -- can be
applied to any description of black hole binaries, including the
effective one-body (EOB) approach \cite{damoureob2001} and extreme
mass ratio inspirals (EMRIs) modeled by the Kerr spacetime \cite{levin2008}.
Additionally, the closed-orbit taxonomy offers a new terrain for the
comparison of the PN expansion to fully relativistic treatments
\cite{{campanelli2008},{hinder2008},{boyle2008}}. Although
quantitative results will change in improved approximations, the
qualitative features should be robust, as the detection of zoom-whirl
behavior in
fully relativistic numerical experiments implies \cite{pretorius2007}.

\section{Preview}

To provide the reader with a road map through intermediate results
accumulated on our way to the periodic taxonomy,
we preview some highlights here.

Our method can be broken into two main steps:

\begin{figure}
\hspace{-155pt}
  \centering
\hfill
\includegraphics[width=90mm]{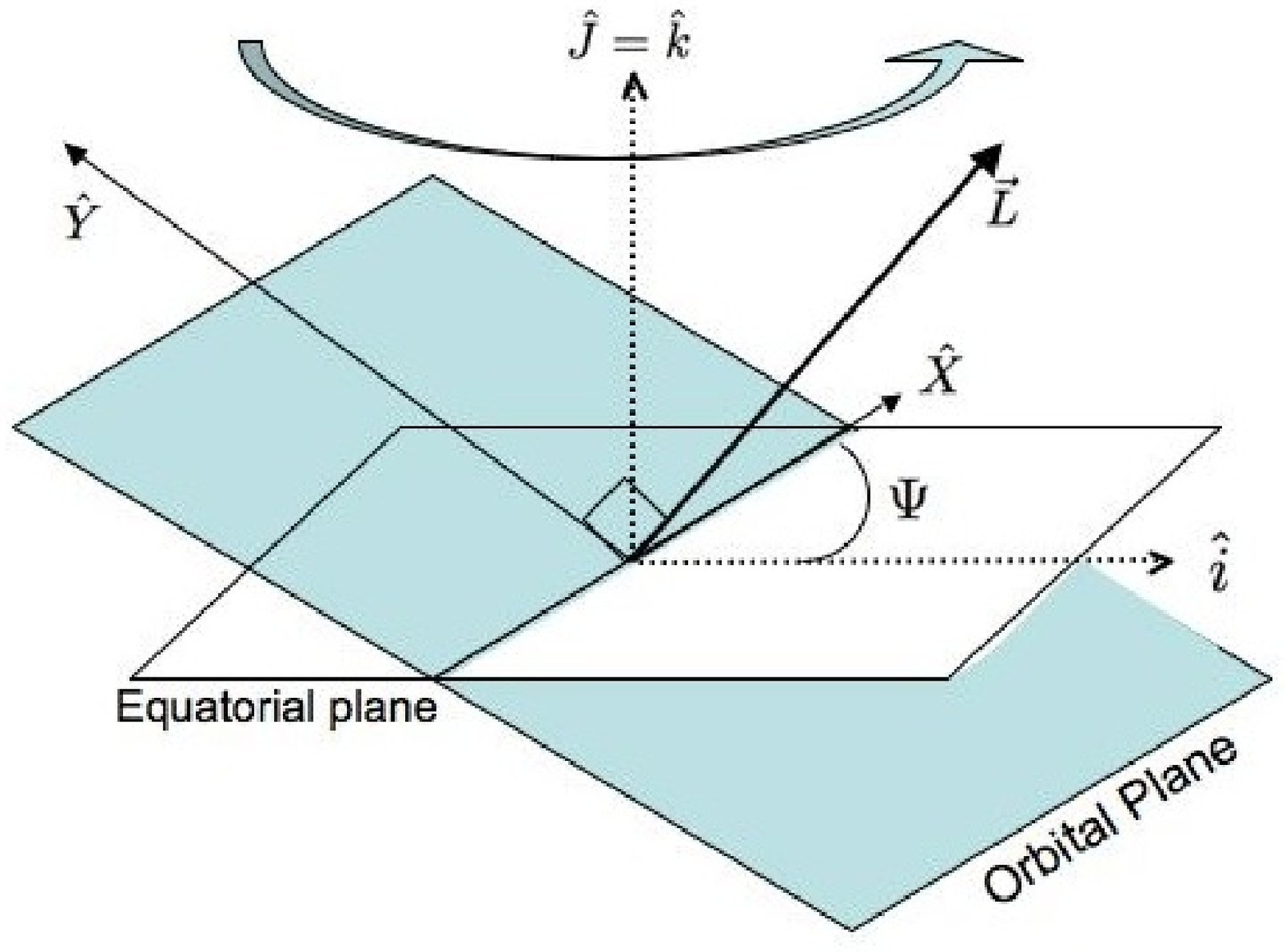}
\hspace{0pt}
\includegraphics[width=60mm]{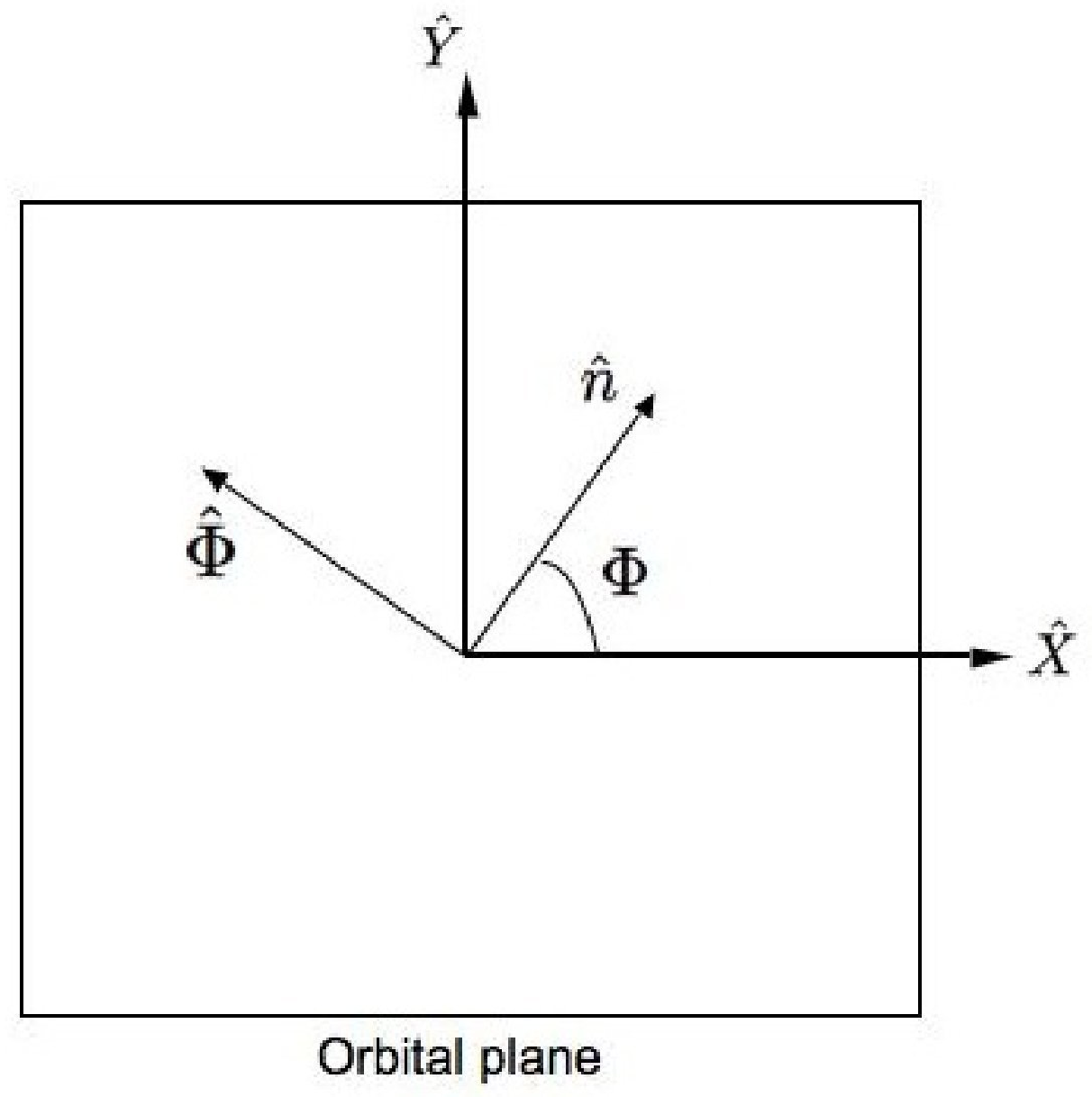}
\hfill
  \caption{Top: The orbital plane precesses around the $\bhj=\bhk$
  axis through the angle $\Psi$.
Bottom: The orbital plane can be spanned by the vectors $(\bhx,\bhy)$
  or the vectors $(\bn,\bhPhi)$.
  \label{orbplane}}
\end{figure}

{\bf 1. Simplified Equations of Motion.}
Since the perihelion precesses and the orbital plane
precesses, the motion around a spinning black hole depends on angles as
well as on the radius. In usual spherical coordinates, the equations
of motion are quite complicated. By working in a non-orthogonal
orbital basis, we show explicitly that the equations of motion 
are independent of (non-orthogonal) angular
variables.

\begin{figure}
  \centering
\includegraphics[width=70mm]{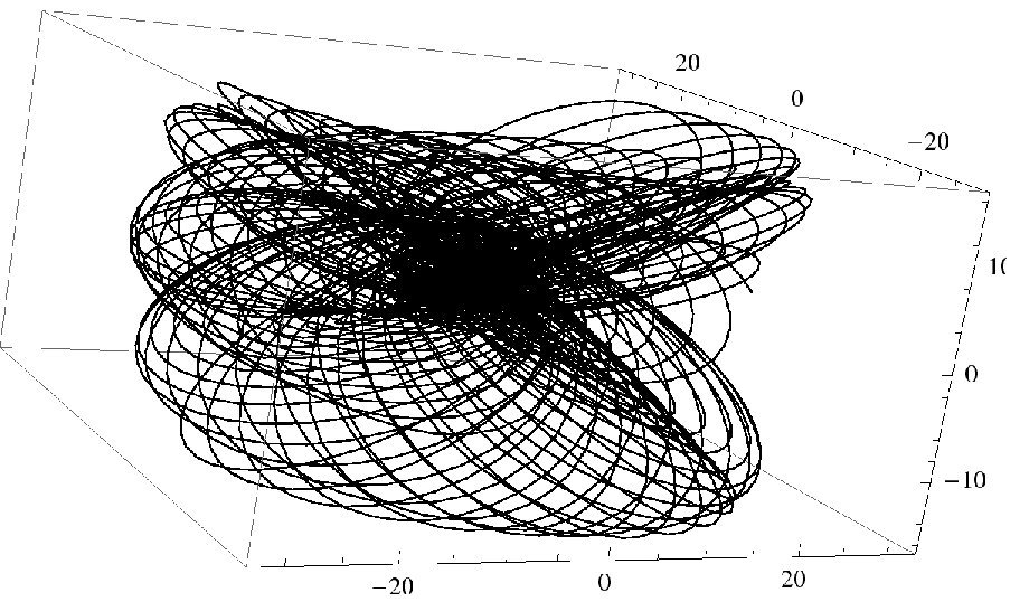}
\hspace{+15pt}
\includegraphics[width=45mm]{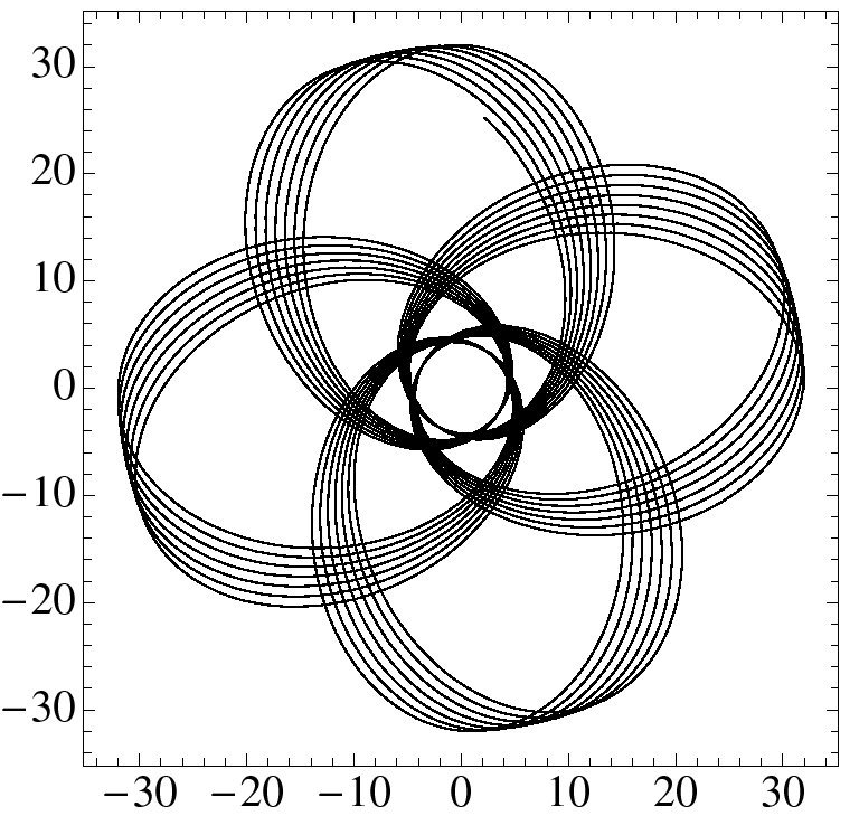}
\hfill
  \caption{
Top:
Fully three-dimensional orbit.
Bottom:
The trajectory as captured in the orbital plane.}
  \label{4leaf}
\end{figure}

Physically, we exploit the observation\footnote{The orbital plane
  is also emphasized in applications of PN dynamics to pulsar
  timing
\cite{{schafer1993},{wex1998},{gong2004},{konigsdorffer2005}}.}
that the orbit lies in the
plane spanned by the coordinate $\br$ and its canonical momentum
$\bp$; that is, the orbital plane is
perpendicular to the orbital angular momentum, $\bl=\br \times \bp$.
The plane itself then precesses around the constant total angular
momentum, $\bj$. The importance of the orbital plane was clear in some
  of the earliest papers on spin-precession \cite{apostolatos1994}, although that
  early work generally imposed a quasi-circular restriction on the orbits. We 
decompose all motion into precession of the perihelion
within the orbital plane with a precession of the entire plane
superimposed, with no restrictions or approximations. A
  preview of the explicit construction is shown in Fig.\
\ref{orbplane}. A fully precessing orbit is shown on the top in Fig.\
\ref{4leaf} while on the bottom the orbital plane traps a much
simplified orbit, reminiscent of the equatorial orbits of Kerr
black holes \cite{levin2008}.

The
simplified Hamilton's equations immediately inform us that 
all eccentric orbits have constant
aphelia and perihelia.\footnote{The constancy of periastron and
  apastron for every orbit might must have been implicitly
understood in Refs.\ \cite{{hartl2005},{damoureob2001}}.} When the aphelia
and perihelia are one and the same, we have non-equatorial constant radius
orbits, also known as
spherical orbits (as previously found in
\cite{{damoureob2001},{buonanno2006}}). The spherical orbits are not
necessarily periodic; they fill out a band on the surface of a sphere.
They are nonetheless significant in our
campaign to fully dissect the dynamics and are treated in a companion
paper
\cite{companion}.

The simplified Hamilton's equations show that zoom-whirl
patterns will be symmetric from one radial cycle to another when
viewed in the orbital plane, as can be seen on the right of Fig.\
\ref{4leaf}. A related subtle feature
is that
the three coordinate velocities in the orbital
basis depend only on radius and are therefore periodic as an orbit
executes a radial cycle from apastron
to apastron.
Taken together these symmetries in the orbital plane
are intriguing for gravitational wave analysis.
The waveforms must be decomposable into the orbital basis and
therefore Fourier decomposable into the three fundamental frequencies
that are the time average over one radial cycle of the instantaneous
velocities.
It remains to be seen how advantageous this might be for gravitational
wave astronomy.

We restrict ourselves to completing the dynamical
picture in this article since it is the dynamics that shapes
the gravitational waves.
In this spirit,
step 1 above allows us to proceed to step 2:

{\bf 2. Taxonomy of Fully 3D Orbits.}
We offer a method to completely
taxonomize the dynamics
with the restriction that only one of the black holes
spins. Our approach includes
{\it all} fully three-dimensional orbits described by the third-order
Post-Newtonian (3PN) Hamiltonian plus spin-orbit couplings. 

Our taxonomy extends the periodic tables for Kerr equatorial
orbits \cite{levin2008} to fully non-equatorial orbits of
comparable mass black hole binaries. In Ref.\ \cite{levin2008}, we
introduced a taxonomy for equatorial Kerr motion with the following
salient features.
Each entry in the Kerr periodic tables of Ref.\ \cite{levin2008} is a perfectly
closed equatorial orbit identified by a rational number
\begin{equation}
q=w+\frac{v}{z}
\end{equation}
where $w$ counts the number of whirls, $z$ counts the number of
leaves, and $v$ indicates the order in which the leaves are traced out.
Since the rationals are
dense on the number line, the periodics are dense in phase
space. Consequently, {\it any} generic equatorial orbit can be {\it arbitrarily}
well-approximated by a nearby periodic orbit. In this way, any generic
orbit is
approximately equivalent to a high-leaf orbit (high
$z$). Additionally, any generic orbit can be approximated as a
precession around a low-leaf orbit, a technique that might ultimately
benefit signal extraction. 

Our ambition in this paper is both to extend the taxonomy to comparable
mass binaries and to resolve the non-equatorial motion of
spinning binaries.
Truly periodic three-dimensional motion follows when the trajectory 
closes in the orbital plane {\it and} the precession of the entire plane
closes simultaneously. Fully closed motion requires two
rationals, each representing a ratio of fundamental
frequencies. 
And although in principle there must exist orbits that are
fully periodic in the three-dimensional motion -- as Poincar\' e argued
\cite{poincare1892} -- our
taxonomy of bound orbits needs only the weaker
condition of periodicity in the orbital plane.
Not every orbit that is
closed in the orbital plane will be closed in the full
three-dimensional space. 
In other words, for the less restrictive condition of orbital plane
periodicity we 
only need one rational ratio of frequencies.
As will be explained in detail in \S
\ref{closed},
the aperiodic orbit of Fig.\ \ref{4leaf} can be approximated as
a precession around a 4-leaf clover in the orbital plane.

Although the PN approximation is poor in the strong-field, the
qualitative results should survive a full relativistic
treatment. Spin-spin couplings will impose additional modulations on
the orbital plane picture but 
since spin-spin couplings are
higher-order in the PN expansion, the expectation has been that their
effect can be treated as a perturbation \cite{buonanno2006}. In a companion paper we
will argue that although a small perturbation, the spin-spin couplings
are responsible for the emergence of chaos around unstable orbits \cite{companion}.

We emphasize that for real spinning astrophysical black holes,
the orbits we resolve in
this paper are not an exotic subset of orbits, but rather are descriptive
of {\it all} bound orbits -- {\it all} non-circular orbits 
are captured in the spectrum of rationals.
There is a long-standing
argument that black hole binaries will circularize by the
time they enter the bandwidth of the gravitational
wave observatories. However this is not possible for spinning black
holes. Circular orbits {\it do not exist} for misaligned spins.
Although spherical orbits do exist, they are destroyed by the
spin-spin effects \cite{companion}. 
What's more, black hole pairs formed in dense clusters are not
expected to circularize by the time of merger and are
expected to be plentiful sources for advanced LIGO \cite{{wen2003}}
While we restrict ourselves to
one spinning black hole and one non-spinning in this paper, the
scenario is both astrophysically possible in its own right and theoretically important
to lay the foundation for the two spinning case with spin-spin
included, a task we return to in a companion paper \cite{companion}

We express Hamilton's 
equations in a non-orthogonal orbital
basis in \S \ref{simple}. 
We discuss the closed orbit taxonomy in \S \ref{closed}.
In \S \ref{periodic}, we show periodic tables for two different black
hole binaries, a comparable mass binary and a
non-spinning extreme mass ratio pair.
Appendix \S \ref{orbitalapp} details the projection of Hamilton's equations onto our
orbital basis.
In
the conclusions, \S \ref{conc},  we discuss the modulations predicted from
spin-spin couplings and those imposed by
spinning both black holes.

\section{Hamilton's Equations of Motion in the Orbital Basis}
\label{simple}

The culmination of this section will be the compact form of the
equations of motion (Eqs.\ (\ref{eoms})) in a non-orthonormal orbital basis. 
To get there will require a few short subsections. We begin
with the 3PN Hamiltonian including spin-orbit couplings.

\subsection{The 3PN Hamiltonian + Spin-Orbit Couplings}
\label{simplesub}

Although the 3PN Hamiltonian is nearly a page long, the Hamiltonian
formulation of black hole pairs has certain advantages
over the Kerr fully relativistic description of test particle motion
around a single black hole. Most notable 
in this context, the ADM-Hamiltonian effectively describes center of
mass motion in flat space. This will allow us to manipulate spatial
vectors at will and locate the orbital plane.(This work has
  suggested a means to generalize to the fully relativistic Kerr
  system. That research is in progress.)

To begin, take 
the 3PN Hamiltonian including spin-orbit coupling as it is
conventionally written in dimensionless coordinates. 
If ${\bf{\cal  R}}$ is the ADM coordinate vector and ${\bf \cal P}$ is
the ADM momentum vector, then
the dimensionless center-of-mass coordinate vector is ${\bf r}={\bf \cal R}/M$ and its canonical
momentum is
 ${\bf p}={\bf \cal P}/\mu$ where
the total mass is
$M=m_1+m_2$ for a pair with black
hole masses $m_1$ and $m_2$ and 
the reduced mass is $\mu=m_1m_2/M$ with the dimensionless combination
$\eta =\mu/M$. All vector quantities
will always be in bold so that $r$ is to be understood as the magnitude 
$r=\sqrt{\br\cdot \br}$. Unit vectors such as $\bn = \br/r$ will additionally
carry a hat as well as being bold.
The dimensionless reduced Hamiltonian $\H={\cal H}/\mu$,
where ${\cal H}$ is the physical Hamiltonian, can be written to 3PN
order as \cite{{schaefer1985},{damour1988},{jaranowski1998},{damourpn2000},{damourpn2000:2},{damourpn2001}}
\begin{widetext}
\begin{equation}
\H=\H_{PN}+\H_{SO} \quad ,
\label{ham1}
\end{equation}
where
\begin{equation}
\H_{PN} = \H_N+\H_{1PN}+\H_{2PN}+\H_{3PN}
\label{hpnlist}
\end{equation}
\begin{align}
\H_N&=\frac{{\bp}^2}{2}-\frac{1}{r} &  & \\
\H_{1PN}&=\frac{1}{8}\left (3\eta-1\right ) \left ({\bp}^2\right )^2
-\frac{1}{2}\left [\left (3+\eta\right ){\bp}^2+\eta({\bn} \cdot 
{\bf  p})^2\right ] \frac{1}{r}+\frac{1}{2r^2} &\nonumber \\
\H_{2PN}&= \frac{1}{16} \left (1-5\eta+5\eta^2\right ) \left ({\bp}^2\right )^3
+\frac{1}{8}\left [\left (5-20\eta-3\eta^2\right)\left ({\bp}^2\right )^2 \right.&\nonumber \\
&\left.- 2\eta^2({\bn} \cdot {\bp})^2{\bp}^2
-3\eta^2({\bn} \cdot {\bp})^4\right ] \frac{1}{r} &\nonumber \\
& +\frac{1}{2}\left [\left (5+8\eta\right ){\bp}^2
+3\eta({\bn} \cdot {\bp})^2\right ] \frac{1}{r^2}
-\frac{1}{4}\left( 1+3\eta \right)\frac{1}{r^3} &
\nonumber \end{align}
\begin{align}
\H_{3PN}&=\frac{1}{128}\left (-5+35\eta-70\eta^2+35\eta^3\right )
\left ({\bp}^2\right )^4
+\frac{1}{16}\left [\left (-7+42\eta-53\eta^2-5\eta^3\right )\left
  ({\bp}^2\right )^3\right. &\nonumber \\
& \left. +(2-3\eta)\eta^2({\bn} \cdot {\bp})^2({\bp}^2)^2
+3(1-\eta)\eta^2({\bn} \cdot {\bp})^4{\bp}^2-5\eta^3({\bn} \cdot
{\bp})^6\right ]\frac{1}{r}&\nonumber \\
&+\left [\frac{1}{16}(-27+136\eta+109\eta^2)({\bp}^2)^2
+\frac{1}{16}(17+30\eta)\eta({\bn} \cdot {\bp})^2{\bp}^2 +
\frac{1}{12}(5+43\eta)\eta({\bn} \cdot {\bp})^4\right ]
\frac{1}{r^2}&
\nonumber \\
&+\left \{ \frac{1}{192}\left [-600+\left
  (3\pi^2- 1340 \right  )
\eta-552\eta^2\right ]{\bp}^2 
-\frac{1}{64}\left(340+3\pi^2+112\eta\right )\eta
({\bn} \cdot {\bp})^2\right \} \frac{1}{r^3}&\nonumber \\
&+\frac{1}{96}\left [ 12+\left (872-63\pi^2\right )
  \eta\right ]\frac{1}{r^4}\quad .&
\label{ahterms}
\end{align}
\end{widetext}
The reduced spin-orbit Hamiltonian is
\begin{equation}
\H_{SO}=\delta_1 \frac{\bl\cdot \bs}{r^3}
\label{HSO}
\end{equation}
where 
in this paper, we restrict our analysis to only one spinning body with reduced spin
\begin{equation}
\bs={\bf a} (m_1^2/\mu M)\quad ,
\end{equation} 
mass $m_1$, and
\begin{equation}
\delta_1\equiv\left (2+\frac{3m_2}{2m_1}\right )\eta \quad .
\end{equation}
Physical values of the dimensionless spin amplitude range over
$0\le a\le 1$. 
In a companion paper, we will generalize to two spinning bodies
\cite{companion}.
We omit
spin-spin coupling terms.
The reduced orbital angular momentum is
\begin{equation}
{\bl}={\br}\times {\bp} \quad .
\label{l}
\end{equation}
Notice that with units included, the physical orbital angular momentum
is $\bl \mu M$.

The equations of motion are given by
\begin{equation}
\dot {\br}=\frac{\partial \H}{\partial {\bp}} \quad ,\quad
\dot{\bp}=-\frac{\partial{\H}}{\partial {\br}}
\label{hameoms}
\end{equation}
and the evolution equation for the spins and the angular momentum can
be found from the Poisson brackets:
\begin{eqnarray}
\dot {\bs} &=& \{{\bs},\H\}=\frac{\partial \H}{\partial {\bs}}\times \bs \nonumber \\
\dot {\bl} &=& \{\bl,\H\}=\frac{\partial \H}{\partial {\bl}}\times \bl
\end{eqnarray}
which comes to
\begin{eqnarray}
\dot {\bs} &=& 
\delta_1 \frac{\bl\times \bs}{r^3}\nonumber \\
\dot {\bf L} &=&\delta_1 \frac{\bs\times \bf L}{r^3}
\quad .
\label{dL}
\end{eqnarray}

\subsection{Conserved Quantities}
\label{conserved}

It is well-known that this system has many useful conserved
quantities.\footnote{
Although it is by now well-confirmed that there is chaos
  when the black holes spin
  \cite{{suzuki1997},{levin2000},{levin2003},{cornish2002},{cornish2003},{hartl2005},{wu2007},{wu2008}},
we are dealing with a restricted situation of only one body spinning
in the Hamiltonian formulation and our orbits are not chaotic to
  this order in the approximation \cite{gopakumar2005}. This
  is not in conflict with earlier work on chaos in
  the Lagrangian approximation \cite{levin2006}.
At higher order including spin-spin
  couplings, a pair of spinning black holes loses
  constants of the motion opening a window for chaotic motion \cite{companion}. }
The Hamiltonian is conserved by construction. 
The conservation of total angular momentum follows from Eqs.\ (\ref{dL})
\begin{equation}
\bj=\bl+\bs \quad .
\end{equation}
Also conserved are the magnitude of $S$ and $L$, as can be confirmed by
taking the dot-products with Eqs. (\ref{dL}):
\begin{align}
\bs\cdot \dot{\bs}&=\half \frac{d}{dt}(S^2)\propto \bl\cdot (\bs \times \bl) =0\nonumber \\
\bl\cdot \dot{\bl}&=\half \frac{d}{dt}(L^2) \propto \bs \cdot (\bl \times
\bs)=0
\end{align}
Finally, 
the component of $\bl$ in the
$\bj$ direction
must be conserved as can be
seen from 
\begin{align}
\dot {\bf L} &=\delta_1 \frac{\bs\times \bf
  L}{r^3} \nonumber \\
&=\delta_1 \frac{\bj\times \bf L}{r^3} \end{align}
from which it follows that the change in $\bl$ is always perpendicular
to $\bj$. 

\subsection{The equations of motion}
\label{eomsection}

We want to express the equations of motion derived from Eq.\
(\ref{hameoms})
in the following form:
\begin{align}
\dot{\br}&=A\bp +B\bn +{\rm spin\ pieces} \nonumber \\
\dot{\bp}&=C\bp +D\bn +{\rm spin\ pieces} 
\quad .
\label{eomform}
\end{align}
This form helps consolidate the equations of motion before we project
from the vector equations to component equations in the next section
(\S \ref{orbital}). We need to identify the functions $A, B, C, D$ in
terms of derivatives on the Hamiltonian, which can be thought of as a function of
$(r,\bp,(\bn\cdot\bp))$.
Considering the non-spinning piece, $\H_{PN}$ first,
we break up the partial derivatives on the right hand side of
Hamilton's equations in the
following way,
\begin{equation}
\left.\frac{\partial \H_{PN}}{\partial {\bp}}\right |_{\br}  =
\left.\frac{\partial \H_{PN}}{\partial
  {\bp^2}}\right|_{r,(\bn\cdot\bp)}\frac{\partial \bp^2}{\partial \bp}+
\left.\frac{\partial \H_{PN}}{\partial
  {(\bn\cdot\bp)}}\right|_{r,\bp}\frac{\partial (\bn\cdot \bp)}{\partial \bp}
\nonumber
\end{equation}
and
\begin{equation}
-\left. \frac{\partial \H_{PN}}{\partial {\br}}\right |_{\bp} =
-\left.\frac{\partial \H_{PN}}{\partial (\bn\cdot \bp)}\right |_{r,\bp}
\frac{\partial (\bn\cdot \bp)}{\partial
  \bn}\frac{\partial \bn}{\partial \br}
-\left. \frac{\partial {\H_{PN}}}{\partial r}\right |_{\bp,(\bn\cdot\bp)}
\frac{\partial r}{\partial\br}  ,
\nonumber
\end{equation}
where we are careful to indicate the quantities held fixed in each term.
Using
\begin{equation}
\frac{\partial(\bn\cdot \bp)}{\partial \bn}\frac{\partial
  \bn}{\partial \br} = \frac{\bp}{r}-\frac{(\bn\cdot
  \bp)}{r}{\bn}
\end{equation}
we define
\begin{align}
A& \equiv 2\left.\frac{\partial \H_{PN}}{\partial \bp^2} \right |_{r,(\bn\cdot\bp)}\\
B& \equiv \left.\frac{\partial \H_{PN}}{\partial (\bn\cdot \bp)}\right |_{r,\bp} \nonumber \\
C&\equiv  -\frac{1}{r}\left.\frac{\partial \H_{PN}}{\partial (\bn\cdot \bp)}\right |_{r,\bp} =-\frac{B}{r}\nonumber \\
D&\equiv -\left.\frac{\partial \H_{PN}}{\partial r}\right |_{\bp,(\bn\cdot\bp)}+\left.\frac{\partial \H_{PN}}{\partial
  (\bn\cdot\bp)}\right |_{r,\bp}\frac{(\bn \cdot \bp)}{r}\nonumber \\
&=-\left.\frac{\partial \H_{PN}}{\partial r}\right |_{\bp,(\bn\cdot\bp)}
-(\bn \cdot \bp)C \quad .
\label{abcd}
\end{align}
The variations of the spinning piece of the Hamiltonian are simply
\begin{eqnarray}
\frac{\partial \H_{SO}}{\partial \bp}&=&\delta_1 \frac{\bs\times
  \br}{r^3} \nonumber \\
-\frac{\partial \H_{SO}}{\partial \br}&=&-\delta_1 \frac{\bp\times \bs}{r^3} 
+3\delta_1\frac{\bl\cdot \bs}{r^4} \bn \quad .
\end{eqnarray}
With these definitions, and making use of $-\bp\times \bs=\bs\times
\bp$, we can write the vector equations of motion
compactly as
\begin{align}
\dot{\br}&=A\bp +B\bn +\delta_1 \frac{\bs\times \br}{r^3} \nonumber \\
\dot{\bp}&=C\bp +D\bn + \delta_1 \frac{\bs\times \bp}{r^3} 
+3\delta_1\frac{\bl\cdot \bs}{r^4} \bn
\quad .
\label{eomshort}
\end{align}
To go from these vector equations to component form requires we choose
a basis. In the next section we will build the orbital basis of Fig.\
\ref{orbplane}, and cast Eqs.\
(\ref{eomshort}) in component form.

\subsection{The Orbital Basis}
\label{orbital}

The clarity of the form of the equations of motion depends on the
basis used to express them. There are several choices although
the 
one we are calling the orbital basis leads to profound clarity of
expression. 
We will build a
{\it non}-orthogonal, unit normalized
basis $(\bn,\bhPhi,\bhPsi)$ in this section.

There are two special planes to consider when spin precession drives
three-dimensional orbits. There is the orbital plane, which is the
plane perpendicular to $\bl $, and there is the equatorial plane,
which is the plane perpendicular to $\bj$ (see Fig.\ \ref{orbplane}).
We will find an orthonormal basis that spans the orbital plane and
then add the motion of the plane itself, which will be in a direction
that is not orthogonal to the orbital plane.
The technique of moving into an orbital plane and projecting
equations of motion onto this basis is familiar from celestial
mechanics and has seen application in the PN approximation to binary
pulsars
\cite{{schafer1993},{wex1998},{gong2004},{konigsdorffer2005}}. We
depart from the usual approach by adopting a non-orthogonal
basis.

The vectors $\bn$ and $\bp$ lie in the orbital plane by the definition
$\bl=r\bn\times\bp$.
We can also span the
orbital plane by orthonormal vectors $(\bn, \bhPhi)$ where
\begin{eqnarray}
\bhPhi &=& \bhl\times \bn \quad .
\end{eqnarray}
We will work in terms of $(\bn,\bhPhi)$ when considering motion in the
orbital plane.

To separate out the precession of the orbital plane from the
three-dimensional motion,
another basis will be useful for intermediate steps. The $(\bhx,\bhy)$ basis
spans the orbital plane but rotates with the
precession of $\bl$:
\begin{align}
\bhx &= \frac{\bhj\times \bhl}{\sin\theta_L} \nonumber \\
\bhy &=\bhl\times \bhx \quad ,
\end{align}
where $\theta_L=\arccos({\bhl\cdot\bhj})$ is constant. 
The $\bhx$ axis is orthogonal to both $\bhj$ and $\bhl$ by
construction and so lies on the intersection of the orbital plane and the equatorial
plane. 
The entire orbital plane maintains a fixed angle $\theta_Y =\pi/2-\theta_L$ 
with $\bj$ as it precesses with the precession of $\bl$.
The motion of the orbital plane can be
understood as the motion of $\bhx$ in the equatorial plane through an
angle $\Psi$
where 
\begin{equation}
\bhPsi=\bhj\times \bhx \quad .
\end{equation} 
Let $\bhj=\bhk$
and let the unit vectors $\bhi$ and $\bhjy$ span the equatorial plane. 
The speed of $\Psi$ motion
can then be determined:
\begin{equation}
\dot {\cos\Psi }={\bf \dot{\bhx}}\cdot \bhi
=
\frac{(\bhj\times \dot{\bl})\cdot \bhi}{L\sin\theta_L}
= \Omega_L \bhPsi\cdot \bhi 
=-\Omega_L\sin\Psi,
\end{equation}
where we have explicitly used the constancy of the magnitude $L$. 
From this we conclude 
\begin{equation}
\Omega_L=
\dot \Psi=\delta_1
\frac{J}{r^3} \ \ .
\label{omegaL}
\end{equation}
does not depend on any angles. 

To find our simplified 
equations of motion we work in the non-orthogonal, unit normalized basis
$(\bn,\bhPhi,\bhPsi)$
in the next section.

\subsection{Final equations of motion in the orbital plane}
\label{finaleoms}

The four equations of motion in the orbital plane are obtained by
projecting Hamilton's equations onto the basis vectors $(\bn,\bhPhi)$. We do this
explicitly in appendix \S \ref{orbitalapp}, where
the projections onto the orbital basis vectors generate the four equations,
\begin{align}
\dot{\br}\cdot \bn&=\frac{\partial \H}{\partial {\bp}} \cdot \bn \nonumber \\
\dot{\br}\cdot \bhPhi&=\frac{\partial \H}{\partial {\bp}} \cdot \bhPhi\nonumber \\
\dot{\bp}\cdot \bn&= -\frac{\partial{\H}}{\partial {\br}}\cdot \bn \nonumber \\
\dot{\bp}\cdot \bhPhi&= -\frac{\partial{\H}}{\partial {\br}}\cdot \bhPhi
\quad .
\label{eomshort2pre}
\end{align}
Compiling the equations of appendix \S \ref{orbitalapp} concisely
gives the remarkably simple equations of motion in the orbital plane
coordinates $(r,\Phi)$ and 
and their canonical momenta $(P_r,P_\Phi)$,
\begin{widetext}

\begin{empheq}[box=\fbox]{alignat=2}
\label{eoms}
\dot r& =AP_r +B\,,\qquad& 
\dot P_r& =A\frac{L^2}{r^3} -\frac{B}{r} P_r +D+3\delta_1\frac{\bs\cdot \bl}{r^4}\\ \nonumber
\dot {\Phi}& =A\frac{L}{r^2}-\frac{L}{J}\Omega_{L}\,,\qquad&
\dot {P}_\Phi& =0
\end{empheq}
\end{widetext}
where $P_\Phi=L$.
The orbital plane precesses at a rate that depends only on $r$, as calculated in the
previous section (\S \ref{orbital}):
\begin{align}
\dot \Psi &=\Omega_L =\delta_1
\frac{J}{r^3} \ \ ,  \quad \dot P_\Psi=0
\label{relates}
\end{align}
with $\delta_1\equiv \left (2+\frac{3m_2}{2m_1}\right )\eta $ and $P_\Psi=L_z$.
For completeness, we can track the precessions of the spin and the
angular momentum:
\begin{eqnarray}
\dot {\bs} &=& 
\Omega_L {\bhj\times \bs}\nonumber \\
\dot {\bf L} &=&\Omega_L{\bhj\times \bf L}
\quad .
\label{dL2}
\end{eqnarray}
As noted earlier, $\bs\cdot\bl=$constant. 
Notice, this basis is explicitly constructed for $\bs \times \bl \ne
0$. When the spin and orbital angular momentum are aligned, anti-aligned, or spin
is zero, then motion is confined to a plane and we should use the usual
equatorial planar basis. This is done explicitly in \S \ref{equatorialbasis}.

The functions $A,B,C,D$ of Eqs.\ (\ref{abcd}) depend only on
$r,P_r$ and constants. 
This can be seen by noting that $A,B,C,D$ are functions of
$(r,(\bn\cdot\bp),\bp^2)$. Now, $\bp$ can be written in terms of
a piece in the radial direction and a piece perpendicular to the
radial direction:
\begin{align}
\bp&=(\bp\cdot \bn)\bn +
(\bn\times \bp)\times \bn\nonumber \\
 &= P_r\bn +\frac{\bl}{r}\times \bn
\end{align}
so that
\begin{equation}
\bp^2=P_r^2+\frac{L^2}{r^2}
\label{decomp}
\end{equation}
and $L$ equals a constant. 
The term $\bp^2$ can therefore be expressed as a function of $(r,P_r)$
only. Meanwhile, $(\bn\cdot \bp)=P_r$ and so
any function of
$(r,(\bn\cdot\bp),\bp^2)$ is equally well a function only of $(r,P_r)$
and constants. 

Consequently,
the above equations of motion
are, amazingly enough, independent\footnote{The simplicity of the
  orbital equations of motion can of course be
  recast in terms of symmetries.
Rotations about $\bhl$ through the angle $\Phi$,
$R_{\bhl}(\Phi)$,
and rotations about $\bhj$ through the angle $\Psi$,
$R_{\bhj}(\Psi)$,
leave the dynamics invariant. The coordinates $\Phi$ and $\Psi $ are cyclic
and $\dot P_\Phi=\dot P_\Psi=0$. The symmetries correspond to
conserved $L$ and $L_z$ with $P_\Phi=L$ and $P_\Psi=L_z$.}
 of the angles $(\Phi,\Psi)$.
The purely radial dependence of the equations of motion 
immediately informs us of four crucial facts (valid to this order in
the PN approximation):
\par
\indent {\bf 1.} { \bf \it There are constant radius orbits.} 
These were already found in Ref.\ \cite{{damoureob2001},{buonanno2006}}
and drop out particularly simply in the orbital basis. In the equatorial plane
these are of course the usual circular obits. Out of the equatorial
plane they have been called spherical orbits in the literature since the orbits trace
out an annulus on the surface of a sphere. We will continue in this
spirit and call them spherical orbits, or, more exactly, constant
radius orbits. 
To see that they exist, notice that the spherical obits correspond to
solutions of $\dot r=\dot P_r=0$ and these roots -- according to Eqs.\
(\ref{eoms}) --
can only depend on constants $(m_1/m_2,S,L,\theta_{LS})$, not on angles. 
In fact, we have a pseudo effective-potential description as we
explain in the next section.

 {\bf 2.} {\bf \it  All eccentric orbits have
constant apastra and perihelia.} Similar to the reasoning above, the
solutions to the condition $P_r=0$ are the turning
points\footnote{When $P_r=0$, $B=0$ and so $\dot
  r=AP_r+B=0$. Therefore we take the $\dot r=0$ condition to be
  synonymous with $P_r=0$.} and thus are the apastron
 and perihelion of any orbit. These values can depend only on constants
\cite{companion}.

\begin{figure}
  \vspace{0pt}
  \centering
\includegraphics[width=60mm]{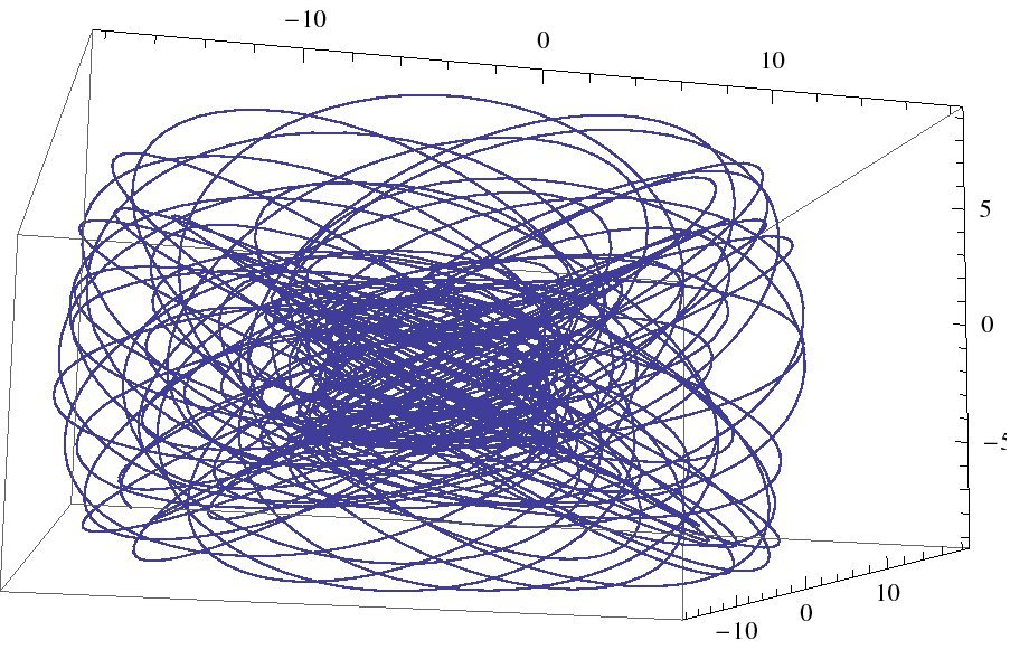}
\hspace{+10pt}
\includegraphics[width=45mm]{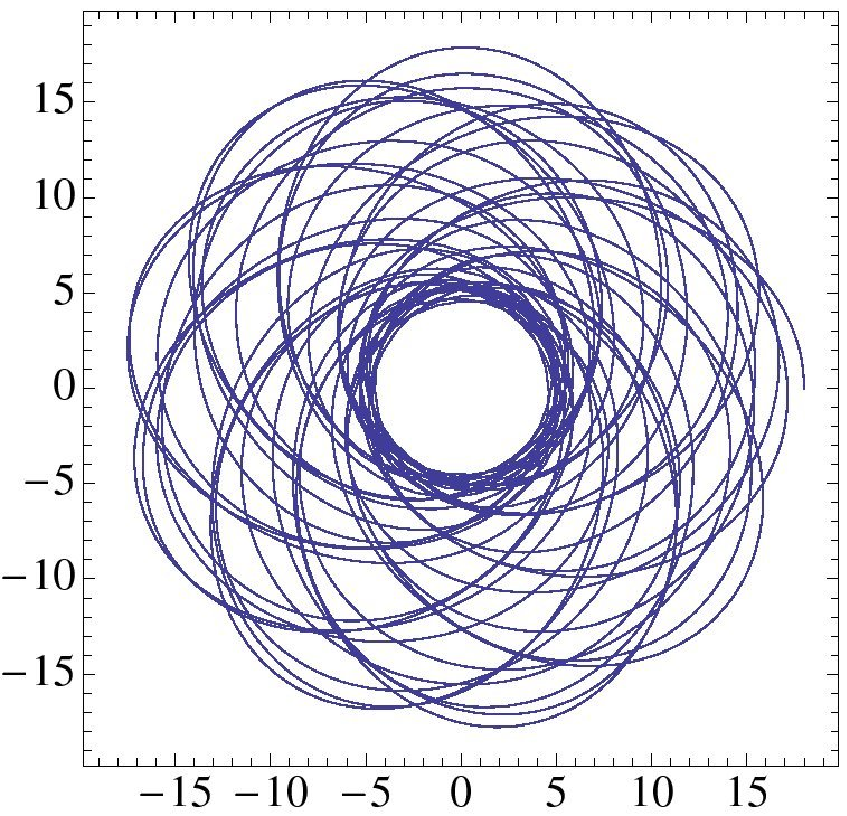}
\hspace{+10pt}
\includegraphics[width=45mm]{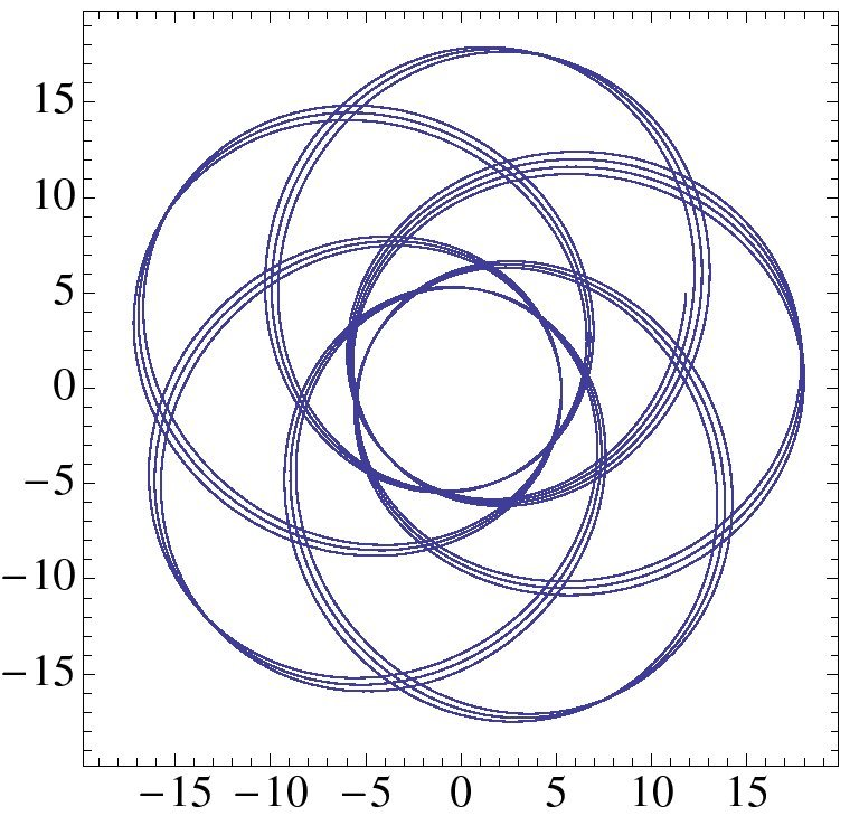}
\hfill
  \caption{A generic orbit in the strong-field.
The initial  conditions are $m_2/m_1=\frac{1}{4}$, $L=3.5$,
$\theta_{LS}  = \frac{\pi}{3}$, $a=0.9$, and $r_i=18$.
Top: The full three-dimensional orbit.  
Middle: A projection of the full orbit onto the equatorial
plane. Looking closely, the angle swept out from leaf to leaf is not
the same under this projection
Bottom: The orbit as caught by the orbital plane. The angle swept out
in the orbital plane from leaf to leaf is always the same.
Further, the constancy of the apastron is clear.
}
  \label{random}
\end{figure}

{\bf 3.} {\bf \it The orbital angle and precessional angle swept out
between successive apastra are constant.} The angles swept out as an
  orbit moves from one apastron to another are simply given by
\begin{align}
\Delta \Phi &= 2\int_{r_a}^{r_p} \frac{\dot \Phi}{\dot r}dr \nonumber \\
\Delta \Psi &= 2\int_{r_a}^{r_p} \frac{\dot \Psi}{\dot r}dr \label{partly}
\end{align}
and depend only on constants.

{\bf 4.} {\bf \it All three coordinates velocities are periodic in
  $r$.} The three coordinate velocities in the orbital basis 
\begin{equation}
\dot r\ , \ \ \dot \Phi \ , \ \ {\rm and} \ \ \dot
\Psi \ ,
\label{freaks}
\end{equation}
depend only on the variable $r$ (or $P_r(r)$) and thereby inherit $r$'s
periodicity. They can be averaged over one radial cycle to define three fundamental
frequencies
\begin{equation}
\omega_r=\frac{2\pi}{T_r} \ , \ \ \omega_\Phi \ , \ \ {\rm and} \ \ \omega_\Psi
\end{equation}
where $T_r$ is the radial period from apastron to apastron (We will
define $\omega_\Phi$ and $\omega_\Psi$ a bit later in Eqs.\
(\ref{phiaverage}) and (\ref{psiaverage})).\footnote{Points
   1 and 2 are also true for Kerr non-equatorial orbits, although 3 and 4 are
   not.
}

\begin{figure}
  \vspace{0pt}
  \centering
\includegraphics[width=60mm]{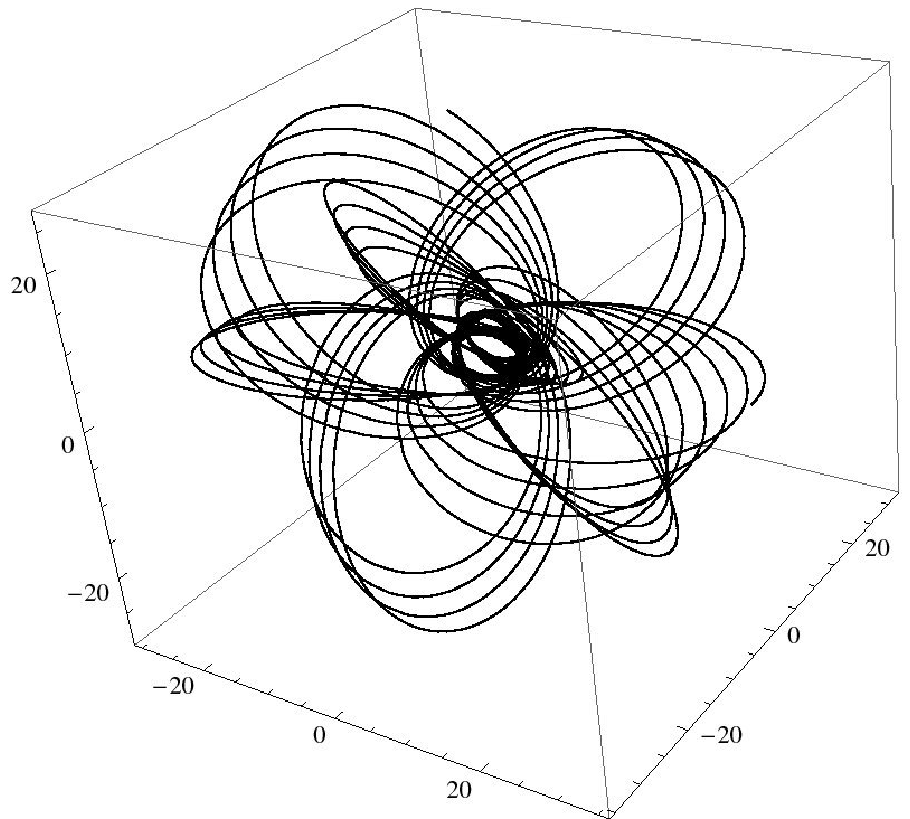}
\hspace{+10pt}
\includegraphics[width=45mm]{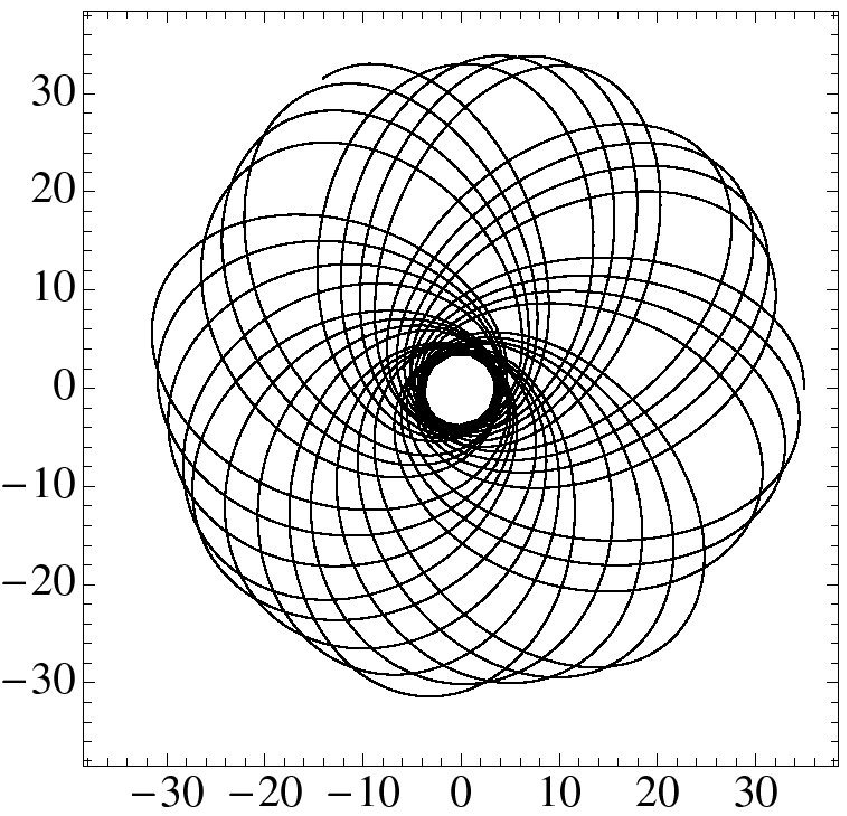}
\hspace{+10pt}
\includegraphics[width=45mm]{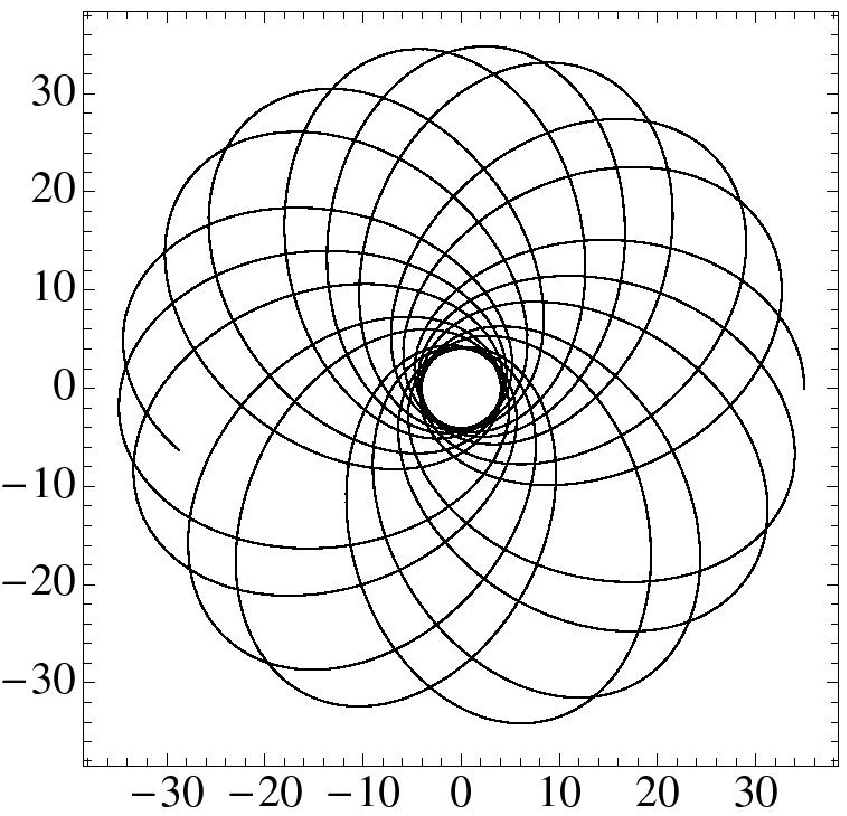}
\hfill
  \caption{The initial conditions are
  $m_2/m_1=\frac{1}{4}$, $L=3.5$, $\theta_{LS} =
  \frac{\pi}{3}$, $a=0.9$
 and 
$E=-0.023548373360051289666$. 
Top: The full orbit. Middle: A
  projection on the equatorial plane. Bottom: The orbit as it appears
  in the orbital plane}
  \label{random2}
\end{figure}

The powerful simplicity of the description in the orbital
basis is visually manifest in Fig.\ \ref{random}. The figure shows a
fully three-dimensional,
generic orbit on the top. The lowest panel is the same
snapshot captured in the orbital plane. Notice how the orbital plane
reveals the constancy of the three-dimensional apastron and perihelion
as claimed in point 2 above. Also notice that the spacing between
leaves is always symmetric in the orbital plane. Said another way, the angle swept out in
the orbital plane between apastra is always the same,
as claimed in point 3. Neither of these features is apparent
from the fully three-dimensional snapshot or from the projection onto
the {\it equatorial} plane shown in the middle view.
Another generic orbit is shown in Fig.\ \ref{random2}.

The four facts above have 
two significant implications:

$\bullet $ A gravitational waveform can be Fourier decomposed in the
three fundamental frequencies of Eq.\ (\ref{freaks}).

$\bullet $ There should be a spectrum of orbits that are 
closed in the orbital plane, and that spectrum must have a
correspondence with the rationals.
We can therefore generalize the Kerr taxonomy of Ref.\
\cite{levin2008} to non-equatorial orbits of comparable mass black hole binaries. 

We complete the dynamical picture by 
moving on to the periodic taxonomy
for black hole binaries.

\section{Closed Orbit Taxonomy}
\label{closed}

Any dynamical study benefits from locating the closed orbits --
orbits that return to their initial values after some finite
period. Poincar\'e was the first to realize that periodic orbits
structure the entire dynamics \cite{poincare1892}. Although a set of
measure zero, the periodic
orbits are dense in phase space. Consequently, any orbit can be approximated as near
some periodic orbit. In that sense, the periodic set forms the
skeleton of the dynamics. What's more, they are all one needs to
know since to arbitrary precision even
an aperiodic generic orbit is arbitrarily close to some periodic
orbit, though possibly one with very high period.

The periodic set corresponds to a spectrum of rational numbers.
That spectrum of rationals shows that the zoom-whirl behavior
known for extreme-mass-ratio binaries is prevalent in the strong-field
regime of comparable binaries as well. Zoom-whirl behavior is
therefore not exotic but rather the norm for the strong-field. The
spectrum of rationals renders the zoom-whirl behavior of any orbit
quantifiable and unambiguous.

Consider the coordinate velocities of Eq.\ (\ref{freaks}).
Taking the time
average of the $\Phi$-frequency over one radial cycle gives the
fundamental frequency
\begin{equation}
\omega_\Phi=\frac{2}{T_r}\int_{r_a}^{r_p}\frac{\dot \Phi}{\dot r}dr
=\frac{\Delta \Phi}{T_r}=\omega_r \frac{\Delta \Phi}{2\pi} \ \ ,
\label{phiaverage}
\end{equation}
so that 
\begin{equation}
\frac{\omega_\Phi}{\omega_r}=\frac{\Delta \Phi}{2\pi} \, .
\label{endstep}
\end{equation}
An orbit that is closed in the orbital plane has rationally related
frequencies
\begin{equation}
\frac{\omega_\Phi}{\omega_r}=1+q_\Phi \ \ .
\end{equation}
By Eq.\ (\ref{endstep}),
we can interpret the rational in terms of the angle swept out from
leaf to leaf in the orbital plane 
\begin{equation}
\frac{\Delta \Phi}{2\pi}=1+q_\Phi  =1+w_\Phi +\frac{v_\Phi}{z_\Phi}\ \
,
\label{qphi}
\end{equation}
where we have written the rational in terms of a triplet of integers,
as can always be done \cite{levin2008}. 
In the equatorial case we know that $\Delta \Phi>2\pi$ 
for all eccentric orbits; all relativistic orbits overshoot as the
famous precession of the perihelion of Mercury attests. For that
reason we have separated out a $1$ from the definition of $q_\Phi$ in
Eq.\ (\ref{qphi}).\footnote{
There is an interesting anomaly that has to be mentioned.
As it happens, in this 3PN approximation, it is possible for
periodics in the orbital plane to {\it under}-shoot $2\pi$; that is,
from one apastron to another $\Delta \Phi< 2\pi$ ($q_\Phi<0$). This
never happens with Kerr equatorial orbits and may just be a
peculiarity in the approximation, although 
we won't know for certain until the Kerr
non-equatorial case is completed.
Also notice that while some
orbits may undershoot in the orbital plane,
$\Phi$ is not the whole story (see the footnote in \S
\ref{equatorialbasis}) and some of the apparent regression is more
than compensated for by $\Psi$.}

\begin{figure}
  \centering
\includegraphics[width=70mm]{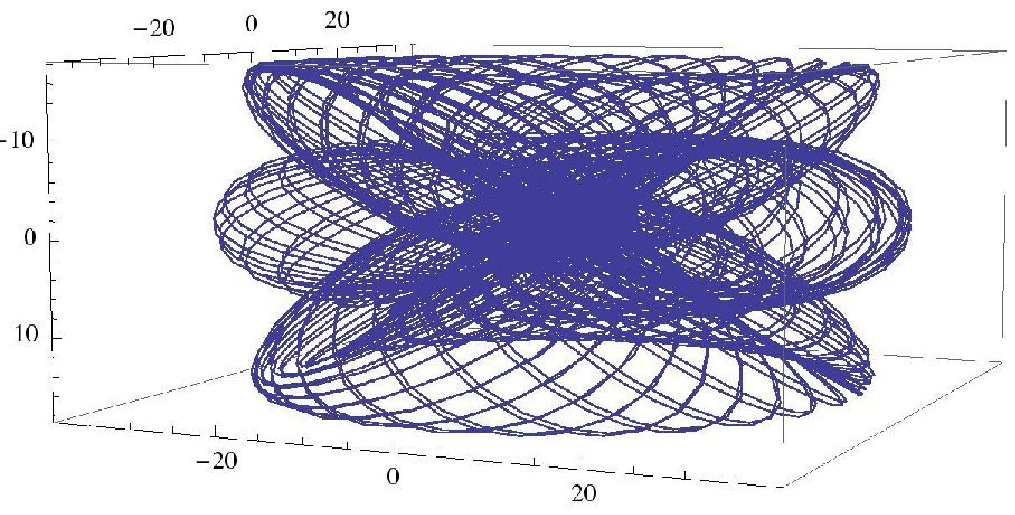}
\hspace{+15pt}
\includegraphics[width=45mm]{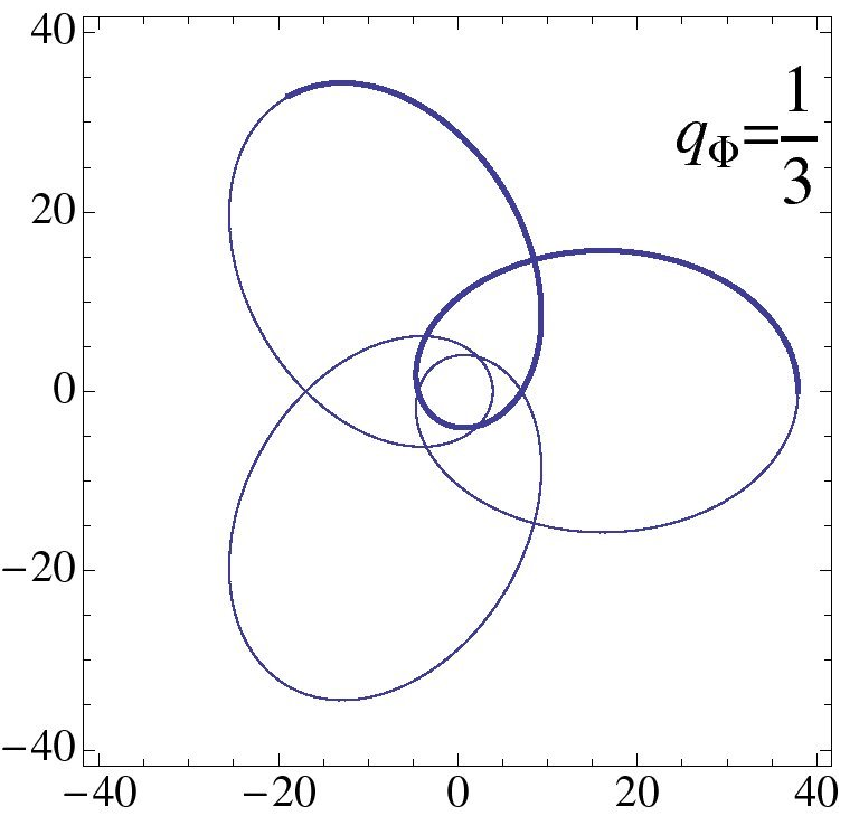}
\hfill
  \caption{A $q_\Phi=1/3$ orbit.
The initial
  conditions are $m_2/m_1=\frac{1}{4}$, $L=3.5$, $\theta_{LS}
  = \frac{\pi}{3}$, $a=0.9$, 
and $E=-0.0220582156$.
Top:
Fully three-dimensional orbit.
Bottom:
The trajectory is a 3-leaf periodic in the orbital plane. The first
radial cycle is in bold.}
  \label{per}
\end{figure}

\begin{figure}
  \centering
\includegraphics[width=70mm]{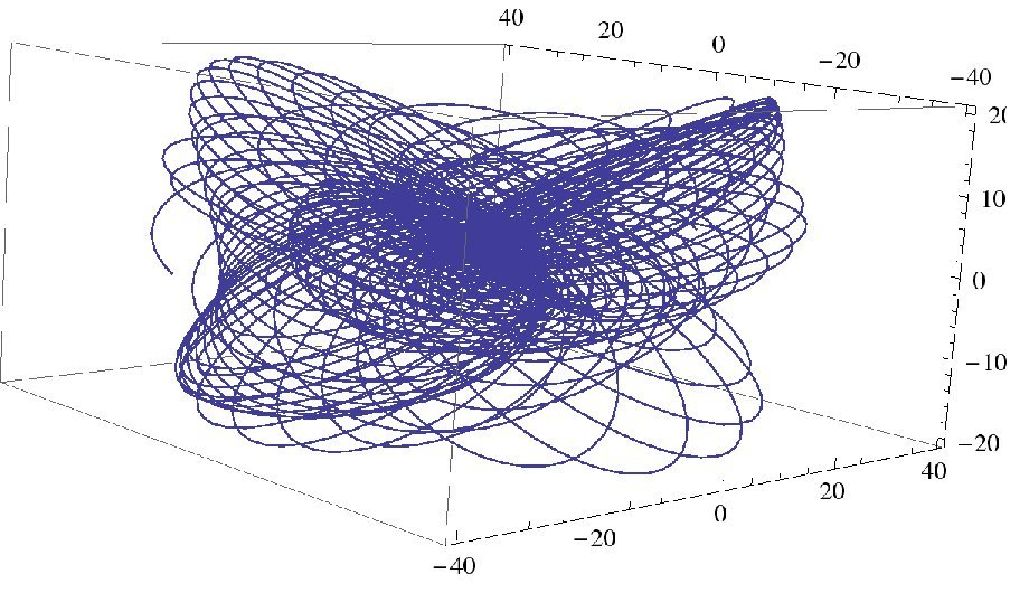}
\hspace{+15pt}
\includegraphics[width=45mm]{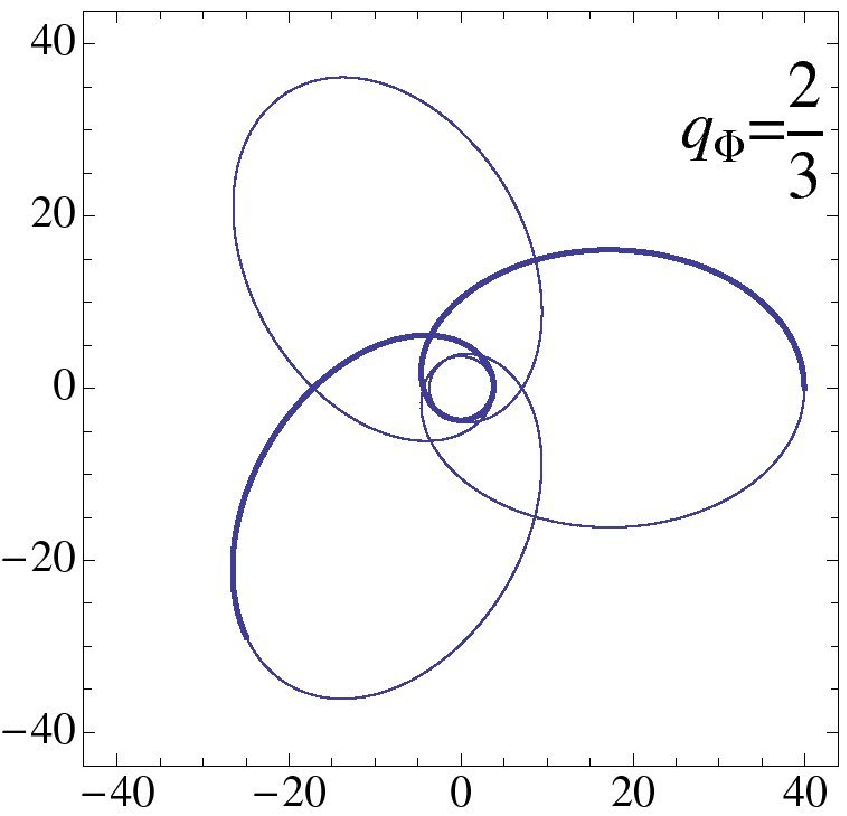}
\hfill
  \caption{A $q_\Phi=2/3$ orbit.
The initial
  conditions are $m_2/m_1=\frac{1}{4}$, $L=3.5$, $\theta_{LS}
  = \frac{\pi}{3}$, $a=0.9$, 
and $E=-0.0211669686$. 
Top:
Fully three-dimensional orbit.
Bottom:
The trajectory is a 3-leaf periodic in the orbital plane
 that skips a leaf each radial
cycle. The first
radial cycle is in bold.
}
  \label{per2}
\end{figure}

By analogy with the equatorial Kerr
case of Ref.\ \cite{levin2008}, $z_\Phi$ counts the number of leaves (or
zooms), $v_\Phi$ specifies the order in which the leaves are traced
out, and $w_\Phi$ counts the number of additional full $2\pi$ whirls
taken between apastron and apastron. 
To clarify the role of $v_\Phi$,
label the leaves sequentially $0$ through $z-1$ starting with the
initial apastra. Then $v_\Phi$ equals the number of the leaf that the
orbit jumps to after
the starting apastron.
The meaning of the rational is best illustrated with an example.
An orbit with $q_\Phi=1/3 $ 
is shown in Fig.\ \ref{per}. This is a 3-leaf orbit $(z_\Phi=3)$ that moves to the
first leaf in the pattern $(v_\Phi=1)$. Since $w_\Phi=0$, there are no additional
whirls from leaf to leaf. In Fig.\ \ref{per2}, we show a $q_\Phi=2/3$
orbit. That is, a
 3-leaf orbit $(z_\Phi=3)$ that moves to the
second leaf in the pattern $(v_\Phi=2)$. Since $w_\Phi=0$ there are no additional
whirls from leaf to leaf. 
For a $z_\Phi$-leaf orbit, the range of $v_\Phi$ for orbits that
overshoot, that is precess, is
\begin{align}
 1 \le v_\Phi & \le z_\Phi-1 \, , \,   &{\rm if} \, z_\Phi>1 &\nonumber \\
v_\Phi &=0 \, , \,  &{\rm if} \, z_\Phi=1 & \quad .
\label{over}
\end{align}
To avoid degeneracy, we require that $z_\Phi$ and $v_\Phi$ be relatively
prime, or in other words, that $q_\Phi=2/4$ is the same as $q_\Phi=1/2$.

The periodic orbits in the orbital plane are a set of measure zero in the space of orbits,
just as the rationals are a set of measure zero in the set of the
real numbers. However, just as the rationals are dense on $\mathbb{R}$, the
periodics are dense in the space of orbits and
so any generic orbit can be arbitrarily well approximated by an orbit
that is closed
in the orbital plane.

\begin{figure}
  \vspace{0pt}
  \centering
\includegraphics[width=70mm]{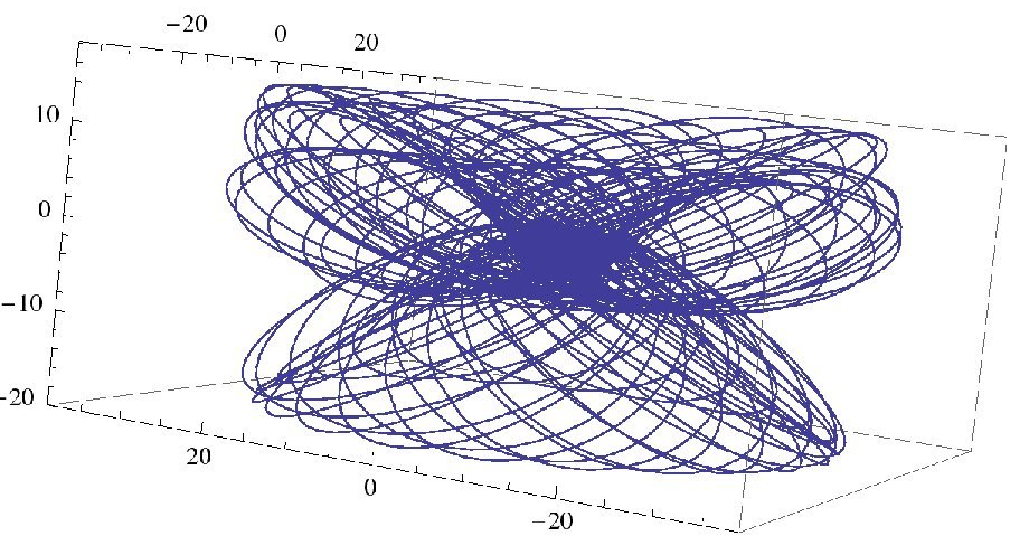}
\hspace{+15pt}
\includegraphics[width=45mm]{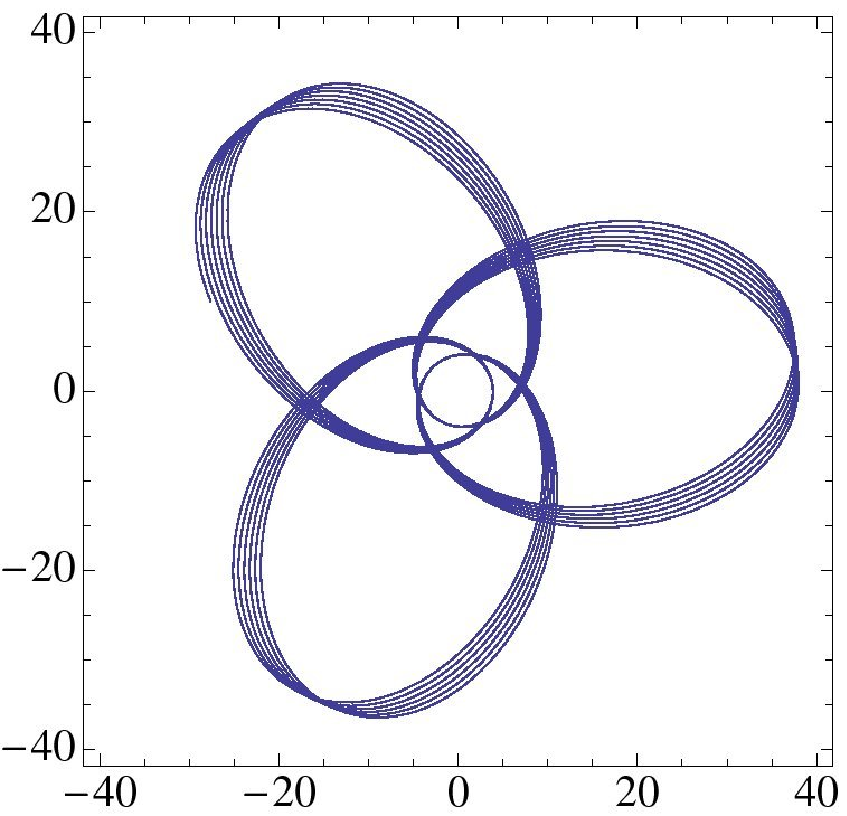}
\hfill
  \caption{An orbit for which $q_\Phi=\frac{67}{200}$.
  The initial conditions are $m_2/m_1=\frac{1}{4}$, $L=3.5$,
  $\theta_{LS} = \frac{\pi}{3}$, $a=0.9$,
and $E=-0.0220323426$
Top: The full three-dimensional orbit.
Bottom: The orbit in the orbital plane is a precession of the exact
$q_\Phi=1/3$ orbit.}
  \label{pernear2}
\end{figure}

For instance, the orbit of Fig.\ \ref{pernear2} is very near the
3-leaf orbit and can be interpreted as a precessing 3-leaf orbit. Or
we could do better by approximating this orbit as a $q_\Phi=67/200$,
that is, an orbit with 200 leaves that skips to the 67th successive leaf in the
pattern with each radial cycle.

By the same token, the randomly selected orbit of Fig.\ \ref{random}
is very nearly a $q_\Phi=1/5$, that is, a 5-leaf clover, and that of
Fig.\ \ref{random2} is very
nearly a $q_\Phi=7/25$ -- an orbit with 25 leaves that skips to the
7th successive leaf in the pattern with each radial cycle.

It is important to notice that although the orbit of Fig.\ \ref{per}
closes in the orbital plane after 3 radial cycles, it does not close
in the full three-dimensional space since
the orbital plane has not returned to its original location after only 3 radial
cycles. 
A fully closed orbit also has to close in $\Psi$.
Taking the time
average of the rate of change $\dot \Psi$ over one radial cycle
gives the fundamental frequency
\begin{equation}
\omega_\Psi=\frac{2}{T_r}\int_{r_a}^{r_p}\frac{\dot \Psi}{\dot r}dr
=\frac{\Delta \Psi}{T_r}=\omega_r \frac{\Delta \Psi}{2\pi} \ \ ,
\label{psiaverage}
\end{equation}
so that 
\begin{equation}
\frac{\omega_\Psi}{\omega_r}=\frac{\Delta \Psi}{2\pi} \, .
\label{endstep2}
\end{equation}
The average precessional frequency may not be rationally related to
the radial frequency for a rational $q_\Phi$: 
\begin{equation}
\frac{\omega_\Psi}{\omega_r}=\sigma_\Psi \ \ ,
\end{equation}
where by $\sigma_\Psi $ we mean any real number, not just a rational.
This time we do not separate out a 1 from the definition of the
number. So, $\sigma_\Psi$ represents the fraction of $2\pi$ swept out
as the plane precesses. The orbit of Fig.\ \ref{per} has a $\sigma_\Psi\approx
0.346...$, where we have only listed the first 3 significant figures. 
Although numerical imprecision of the computer truncates this at a
finite number of digits, and therefore effectively approximates
$\sigma_\Psi$ by a rational, it is in principle an irrational. 
After 3 radial cycles, the orbit has closed in the orbital plane but
not in three dimensions. The entire orbital plane has overshot its
initial location by $3\sigma_\Psi-1\sim 0.038...$.  Therefore, even the
3-leaf clover in the orbital plane of Fig.\ \ref{per} will fill out
the surface of the three-dimensional picture.

\begin{figure}
  \centering
\includegraphics[width=60mm]{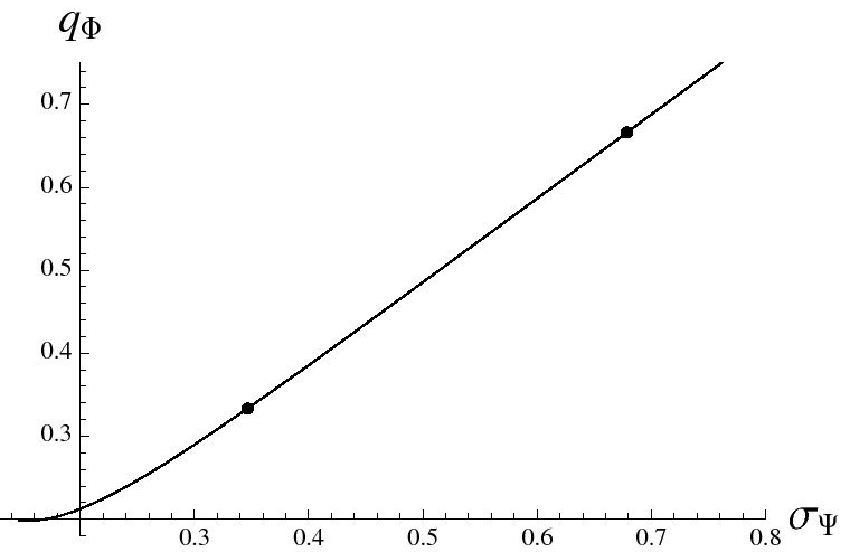}
  \hspace{10pt}
\includegraphics[width=60mm]{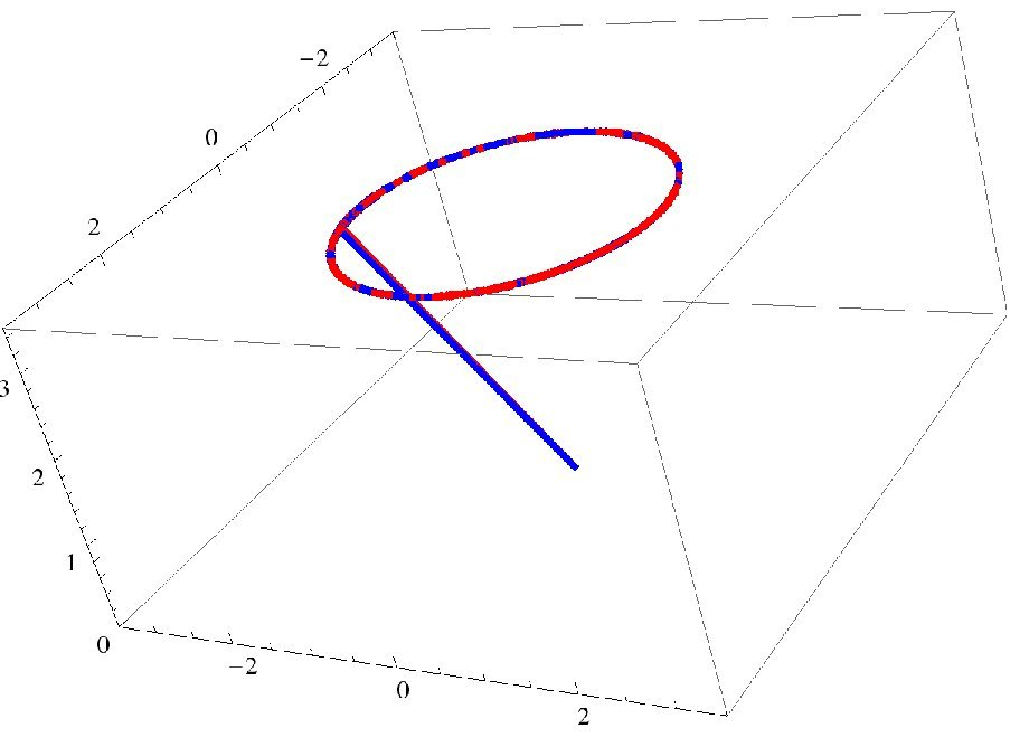}
\hfill
  \caption{$(m_2/m_1=1/4,a=0.9,\theta_{LS}=\pi/3,L=3.5)$. Top:
    $q_\Phi$ versus $\sigma_\Psi$. The dots mark
    $(q_\Phi=1/3,\sigma_\Psi\approx 0.346...)$ and
    $(q_\Phi=2/3,\sigma_\Psi\approx 0.679...)$.
Bottom: The circle
  traced out by the tip of the vector $\bl$ for
  the orbit of Fig.\ \ref{per}. The straight line
  represents the $\bl $ vector when 3 radial cycles have elapsed and
  the $q_\Phi=1/3$ orbit has closed in the orbital plane. The same
  plot for the precessing 3-leaf clover of Fig.\ \ref{pernear2} is
    superposed although the two are so close
  that they cannot be distinguished in the graph. The fact that they
    cannot be distinguished confirms that the two orbits are genuinely
    near each other in 3d as well as in the orbital plane.}
  \label{Lvec}
\end{figure}

In fact, from the equations of motion Eqs.\
(\ref{eoms})-(\ref{relates}),
we know that 
\begin{equation}
1+q_\Phi=-\frac{L}{J}\sigma_\Psi+ f(E,L) \ \ ,
\label{qs}
\end{equation}
where $f(E,L)$ is a function of $E,L$ through the dependence of the
apastra and periastra's dependence on $E,L$.
This curve for
$(m_2/m_1=1/4,a=0.9,\theta_{LS}=\pi/3,L=3.5)$ is shown on the top of
Fig.\ \ref{Lvec}.
Eq.\ (\ref{qs}) is interesting for two reasons. It means that
$\sigma_\Psi $ is not generally a rational number when $q_\Phi$ is
rational. What's even more interesting is that this does not seem to
matter. 
Because $q_\Phi$ effectively fixes the value of $\sigma_\Psi$ for a given $L$ through
relation (\ref{qs}), if a generic orbit is well approximated by a
periodic in the orbital plane, the precession of its orbital plane is
also nearby in the phase space sense.\footnote{Of course, we can alway approximate any irrational, including
$\sigma_\Psi$ by a rational. But then we are describing approximately
periodic orbits as opposed to orbits that are formally exactly
periodic and there doesn't appear to be any advantage in taking this
tack. Hereafter, we will consider $\sigma_\Psi$ to be generally
irrational for any rational $q_\Phi$.}

This
 point is emphasized on the bottom of Fig.\ \ref{Lvec}, which shows the circle
traced out by the precession of the $\bl$ vector for the 3-leaf clover
in the orbital plane of Fig.\ \ref{per}. The straight line from the
origin to the ring indicates the direction of $\bl$ after 3 radial
cycles have elapsed and the 3-leaf clover has executed one complete
period in the orbital plane. The $\sigma_\Psi$ of this orbit is
$\approx 0.346...$ so that in 3 radial cycles the orbital plane has just
overshot its initial location. By comparison, the nearby orbit of Fig.\ \ref{pernear2}
has a $\sigma_\Psi\approx 0.348...$ and its orbital plane similarly
 has just barely 
overshot its initial location. The orbit of Fig.\ \ref{pernear2}
 precesses around the 3-leaf clover of Fig.\ \ref{per},
and its entire orbital plane precesses around $\bj $, sticking close
 to the precession of the
3-leaf clover's
orbital plane. In fact, the precessions of $\bl$ are superposed 
on the bottom of Fig.\ \ref{Lvec} and the
difference between them is imperceptible at the resolution shown.
The fact that the precession of $\bl$ for these two orbits is effectively
 indistinguishable confirms that the two orbits are not only near each
 other in the orbital plane, they are genuinely
near each other in 3D as well.

Formally, the above argument ensures that any orbit can be
approximated as arbitrarily near an orbit that is periodic in the
orbital plane with the same $L$.
In other words, the orbital periodic spectrum for a given $L$ is
dense. If we remove the restriction of comparing orbits of the same
$L$, it follows that the set of orbits periodic in the orbital plane
is dense in the entire space. The argument can be sketched as follows.
According to Poincar\'e, the set of
orbits that is fully periodic in 3D is dense in the phase space. This
set is a subset of the orbital plane periodic set. Therefore, if the
subset is dense, the set itself must be dense.

In short, we can understand the entire
three-dimensional orbital dynamics through orbits that are closed in
the orbital plane and that one rational, not two, is needed for a
taxonomy. These conclusions are of course only valid up to 3PN with
spin-orbit coupling. In the summary we will discuss the modulation 
expected by going to higher order in the approximation.
In the meantime, we move on to the periodic tables for comparable mass binaries.

\section{Periodic Tables}
\label{periodic}

Since {\it every} orbit can be approximated as one that is periodic in the
orbital plane -- just as {\it every} irrational can be approximated by
a rational -- we can build a table of orbits for black hole
binaries with a given mass ratio, spin, and angle between the spin and
the orbital angular momentum $(m_2/m_1,S,L,\theta_{LS})$. Such a
periodic table works in analogy to the chemical periodic table as
illustrated in Fig.\ \ref{equatorial}. Every
entry represents a closed orbit labeled by a rational. The energy and
the rationals both increase monotonically from top to
bottom and then from left to right. Unlike the chemical periodic
table, the black hole periodic tables are infinite since the rationals
are infinite and we show only a handful of entries.

When we discuss specific tables, we will always take all
entries in a given table to have the same $(m_2/m_1,S,L,\theta_{LS})$. The entries vary
only in energy (and therefore in $q_\Phi$). Periodic tables can be
constructed for any values of these parameters. However, some ranges
give fuller tables in the sense that certain ranges permit more
whirliness. To understand the ranges of whirliness requires careful consideration 
of properties of the spherical
orbits. Several results concerning non-equatorial spherical
orbits are presented in \cite{companion}. Here we summarize the pertinent
results for constructing periodic tables with use of a
pseudo effective-potential picture.

\begin{figure}
  \centering
\includegraphics[width=75mm]{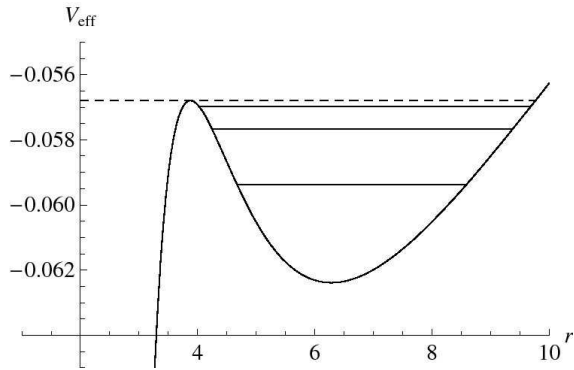}
\hfill
  \caption{$(m_2/m_1=1/4,a=0.5,\theta_{LS}=\pi/4)$. 
The pseudo effective-potential as a function of $r$ shows a maximum at
the unstable spherical orbit and a minimum at the stable
spherical orbit. The homoclinic orbit is indicated with a dashed
line. The other lines correspond to, in ascending order,
$q_\Phi=\frac{2}{5},\frac{1}{2},\frac{2}{3}$. The higher $q_\Phi$
orbits quickly stack together near the
homoclinic orbit.
  \label{extra}}
\end{figure}

A sensible condition for an effective potential formulation is that
the Hamiltonian depend only on $r$ and constants.
Generally, the
Hamiltonian of Eqs.\ (\ref{ham1})-(\ref{ahterms}) does not admit a simple
effective potential formulation since it is a complicated function of
$\bp^2$. We have already argued that $H(r,\bp,\bs)$ can be written as a
radial Hamiltonian $H(r,P_r)$, yet it remains a complicated function
or $P_r$. However, if we only consider 
\begin{equation}
V_{\rm eff}=\H({P_r=0}) \quad ,
\end{equation} 
then we have a good
representation of the effective potential {\it at the turning
 points}. We cannot misuse the $V_{\rm eff}$ by trying to interpret
motion away from the turning points, but it gives a perfectly valid
description of the behavior at aphelia and periastra as well as on
spherical orbits. Hereafter we'll shorthand the term ``pseudo
effective-potential'' by ``effective potential''.
An effective potential for comparable mass binaries is shown in Fig.\ \ref{extra}.

Evident is the lowest energy orbit at the stable spherical
orbit. 
The highest energy non-plunging orbit is the unstable spherical
orbit.
We are interested in energetically 
bound orbits here, i.e. orbits with $E\le0$. 
 If the unstable spherical
orbit has energy $E<0$, then  
the spectrum of periodic orbits densely fills the energy range between the stable and
unstable spherical orbit.
If instead the unstable spherical orbit has energy $E>0$, then the 
the spectrum of periodic orbits densely fills the energy range between
the stable 
spherical orbit
and the marginally bound orbit at $E=0$.
The energy levels of a few periodic
orbits
are indicated by solid lines in Fig.\ \ref{extra}.

Since $q_\Phi$ is monotonic with energy,
the entries in the periodic table are bounded:
\begin{equation}
q_{min} \le q_\Phi \le q_{max} \quad .
\end{equation}
The value of $q_{min}$ is the $q_\Phi$ of the stable spherical orbit
and $q_{max}$ is set by the $q_\Phi$ of the maximum energy bound
orbit. 

The value of $q_{max}$ depends on the largest energy orbit allowed for
that $L$. 
When an unstable spherical orbit exists there is always an
orbit with the same
$E$ and $L$ at a large radius. The maximum of $V_{\rm eff}$ in Fig.\
\ref{extra} marks the unstable spherical orbit. Drawing a line of
constant energy across the effective potential locates the apastron of
the orbit with the same $E$ and $L$ as the unstable spherical trajectory.
When released from apastron, this orbit whirls an infinite
number of times as it approaches the unstable spherical orbit and is
formally a homoclinic orbit; that is, it approaches the same invariant set
in the infinite future and the infinite past.
Although not strictly
periodic -- the homoclinic orbit never returns to apastron -- it can be
considered the infinite winding limit of the 1-leaf periodic orbits
\cite{levin2008}. As such it is the $w_\Phi=\infty$ limit and we
assign homoclinic orbits a $q_\Phi=\infty$.
Consequently, if the range of parameters has a bound unstable spherical
orbit, then it has a ($q_\Phi=\infty$) homoclinic orbit 
and the associated periodic
tables will exhibit much whirliness.

Since, for a given $L$, the stable spherical orbits bound the allowed
energy range from below, they also
bound the value of $q_\Phi$ from below. One might presume that
$q_{min}=0$ but, importantly, this is not the case. 
To see this notice that a spherical orbit obviously does not have a radial
cycle. The $q_{min}$ set by the stable spherical orbit can instead be
thought of as the value of $q_\Phi$ for a nearby eccentric orbit in the
limit that the eccentricity vanishes:
\begin{equation}
q_{min} \rightarrow_{\lim e\rightarrow 0} \frac{\omega_\Phi}{\omega_r}
-1 \quad .
\end{equation}
This allows us to derive the $q_{min}$ from a stability analysis
since the limit of zero eccentricity is effectively the limit of
constant radius, which implies
$\left. \dot \Phi\right |_{r_s} $ is constant and that
$\omega_r=i\lambda_r$ is given by a small perturbation around the
stable spherical orbit.
We then have, just as we found in Ref.\ \cite{levin2008}
that\footnote{
Although most orbits precess in the orbital plane, Eq.\ (\ref{qminc})
actually allows for regression when $\frac{\left.\dot \Phi\right |_{r_s}}{\left
  |{\lambda_r}\right |}<1$.
When the orbital plane and the equatorial plane align $(\bs \times
\bl)=0$, then 
regression seems intuitively obvious. It only means that $\Phi $ is
not the whole story of the motion of $\bn$ and we must also add in
$\Psi $ to see that the actual orbit precesses in the equatorial
plane, as we will do explicitly by moving to the equatorial basis in \S
\ref{equatorialbasis}. We did however also see regression out of the
equatorial plane and it is difficult to say whether it is a flaw in
the PN approximation or if it will survive a full relativistic treatment.
}
\begin{equation}
q_{min}(r_s)=\frac{\left.\dot \Phi\right |_{r_s}}{\left
  |{\lambda_r}\right |}-1
\quad .
\label{qminc}
\end{equation}

We choose for the sake of illustration to consider tables capable of
probing high whirliness. For this reason we stay within
the range set by the last stable spherical orbit (LSSO) and the
homoclinic orbit
$L_{LSSO}<L<L_{homoclinic}$.
Of course, the drawback is that we are pushing the PN
expansion to the breaking point. Although these inner strong-field
orbits probe beyond the confidence of the 3PN approximation,
the general method of constructing periodic tables in a
full relativistic treatment is robust, as proven by the Kerr
demonstration of Ref.\ \cite{levin2008}.

\subsection{Periodic Tables in the Orbital Plane}

\begin{figure}
  \centering
\includegraphics[width=30mm]{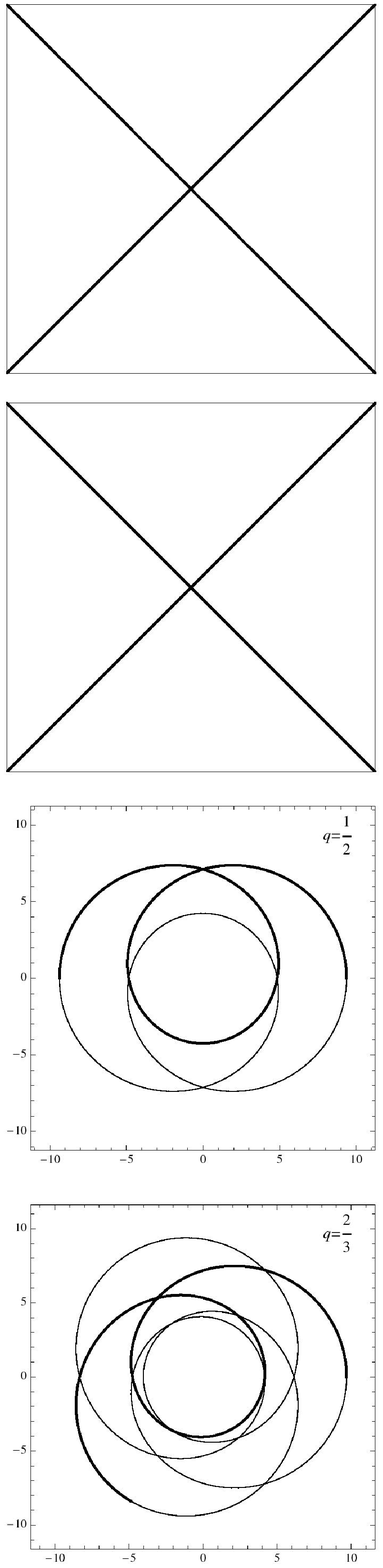}
\hspace{+10pt}
\includegraphics[width=30mm]{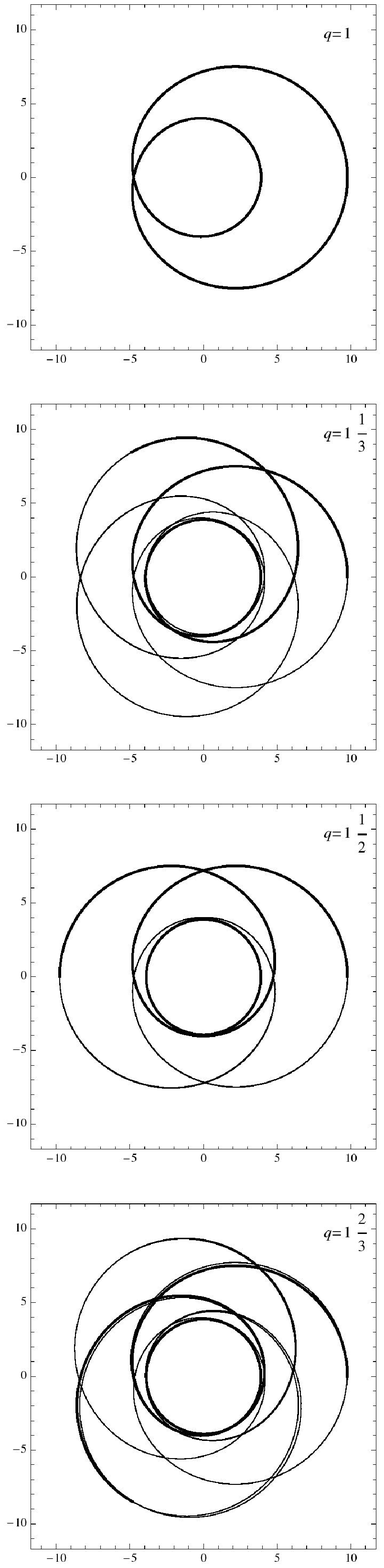}
\hfill
  \caption{A non-equatorial periodic table for which the heavier black
  hole spins with amplitude $a=\frac{1}{2}$, the mass ratio is
  $m_2/m_1=\frac{1}{4}$, $L=3.2$ and $\theta_{LS}=\frac{\pi}{4}$. All valid
  entries up to $z_\Phi=3$ are shown.
The final entry begins
  to show a departure from true periodicity as a result of numerical
  error. The high numerical precision required to keep the simulated
  orbit near a
  perfectly periodic one is a reflection of the tight stacking of high
  $q_\Phi$ orbits near the top of the potential in Fig.\ \ref{extra}.}
  \label{periodicfig}
\end{figure}

The purpose of the Post-Newtonian expansion is of course to
approximate the behavior of comparable mass binaries. A periodic table
for a binary with $m_2/m_1=1/4$ is shown in Fig.\ \ref{periodicfig}.
The heavier black hole has a spin amplitude of $a=1/2$ and the angle
is $\bhl\cdot \bhs = \cos(\pi/4)$. The orbits do not lie in a plane and
are fully three-dimensional, like those of Figs.\ \ref{random} and
\ref{random2}.
Each entry is an orbit that is periodic in the orbital plane,
although not necessarily fully periodic. The energy and the rational
of each entry increase from top to bottom and from left to right. 

Notice that the first two entries are blank before the appearance of
the 2-leaf clover in entry 3. These are blank because, for this
$(m_2/m_1,a,S,L,\theta_{LS})$, the $q_\Phi=0$ and $q_\Phi=1/3$ orbits
simply do not exist 
since
$q_{min}$ is just above $1/3$
(and $q_{max}=\infty$).
We saw this before in the Kerr system \cite{levin2008}. 
The implication is important. {\it All} eccentric orbits -- for this
range of parameters -- show zoom-whirl behavior. None of them look
like the slight precession of the perihelion of Mercury. 

\begin{figure}
  \vspace{0pt}
  \centering
\includegraphics[width=65mm]{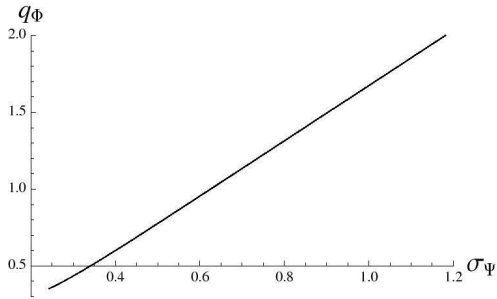}
\hfill
  \caption{$q_\Phi$ versus $\sigma_\Psi$ for the table in
  Fig.\ \ref{periodicfig}.}
  \label{qqplot2}
\end{figure}

{\it Every} orbit in this system can be arbitrarily well-approximated
by an entry in the table, because the
precessional motion of the entire plane is also effectively fixed by
the rational $q_\Phi$ as the plot of Fig.\ \ref{qqplot2} shows. If
$\sigma_\Psi$ could be chosen independently of $q_\Phi$ for a given $L$, our conclusion
would not follow. However, $q_\Phi$ versus $\sigma_\Psi$ lies on a
one-dimensional curve. Once $q_\Phi$ is known, $\sigma_\Psi$ can be read off. Physically
this means that an orbit that is very near a $q_\Phi=1/2$
will precess around the 2-leaf
clover in the orbital plane and that
the precession of the entire orbital plane will be very close to the
precession of the true 2-leaf clover's orbital plane.

\subsection{Periodic Tables in the Equatorial Plane}
\label{equatorialbasis}

Some binary parameters will automatically restrict motion to the {\it
 equatorial} plane and these require special discussion. 
For instance, if neither black hole spins then
the system is spherically symmetric
and all orbits are confined to a plane:
the orbital
plane {\it is} the equatorial plane. Similarly, if one of the black holes
 spins but the spin is aligned with the angular momentum or
 anti-aligned, 
then the motion will again be restricted to the
 equatorial plane. We summarize these three cases as $\bs\times \bl=0$ scenarios.

When $\bs \times \bl =0$, our orbital basis
$(\bn, \bhPhi,\bhPsi)$ is not defined and must be
replaced with the usual planar basis $(\bn, \bhvarphi$),
where $\varphi$ is the usual angle measured\footnote{If we were
  epsilon out of the equatorial plane, we would see regression in the
  orbital plane because much of the motion is taken through
  $\Psi$. Intuitively then, those instance of regression that are just
  barely out of the equatorial plane are not that surprising.}  between $\bn $ and $\bhi$.
The equation of motion for $\varphi $ is simply
\begin{equation}
\dot\varphi=\dot\Phi+\dot \Psi \quad .
\end{equation}
The $q$ we
must use for the equatorial plane of the non-spinning system is then
the same as the one we used in Ref.\ \cite{levin2008}
\begin{equation}
\frac{\omega_\varphi}{\omega_r} =1+q\quad .
\end{equation}
Each entry is specified by
this one rational which represents the ratio of the time averaged orbital angular
frequency in the equatorial plane to the radial frequency, just as in
the Kerr case of Ref.\ \cite{levin2008}. The rational can be read off the
topology of the orbit as
$q=w+v/z$; that is, the number of
whirls, the number of leaves and the order in which the leaves are
laid out fix $q$.

\begin{figure}
  \vspace{+40pt}
  \centering
\hspace{-800pt}

\includegraphics[width=0.15\textwidth]{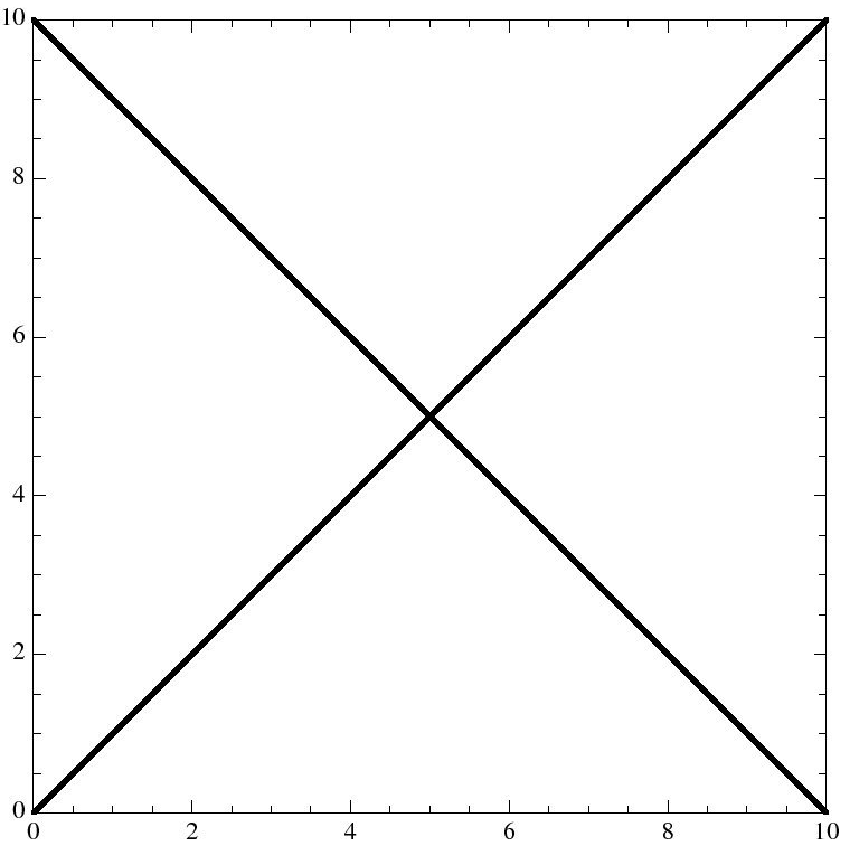}
\includegraphics[width=0.15\textwidth]{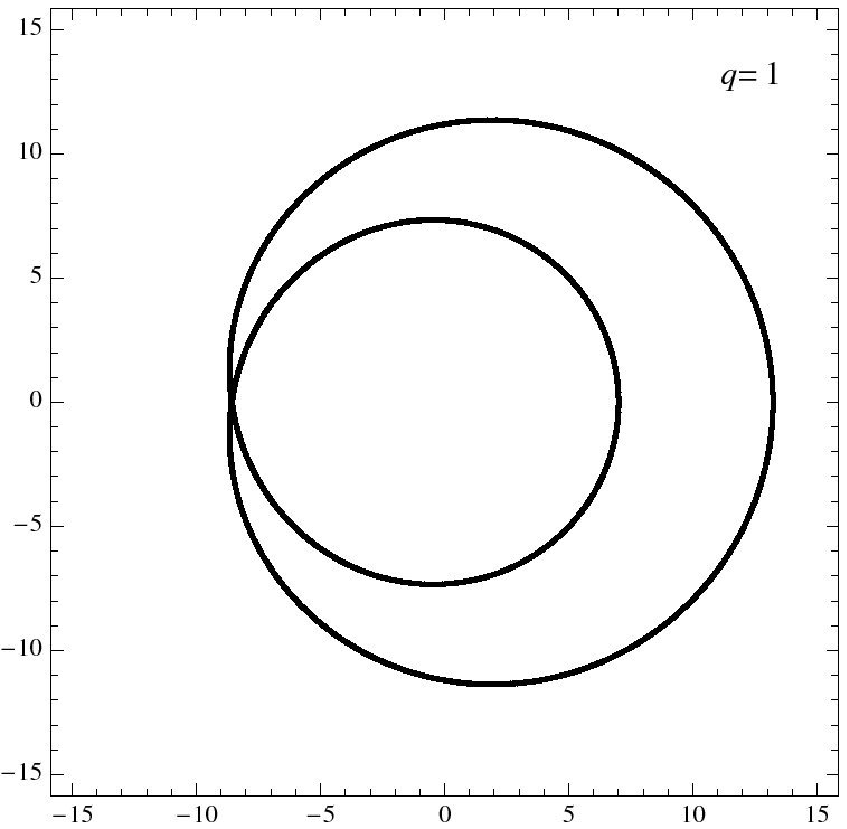}
\includegraphics[width=0.15\textwidth]{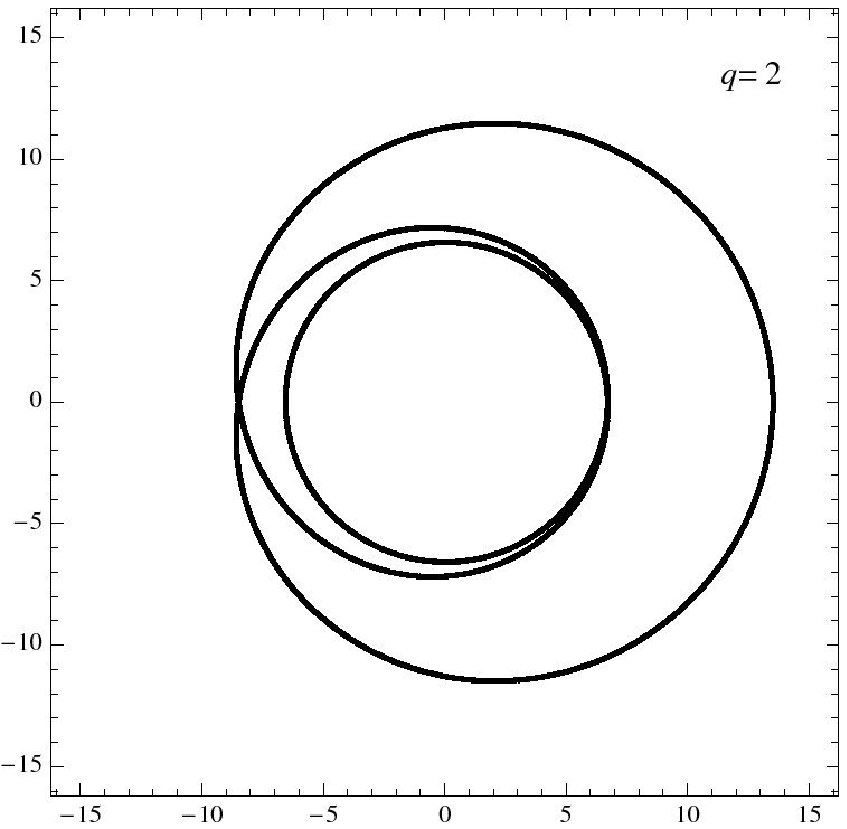}
\hfill


\includegraphics[width=0.15\textwidth]{xplots/000.eps}
\includegraphics[width=0.15\textwidth]{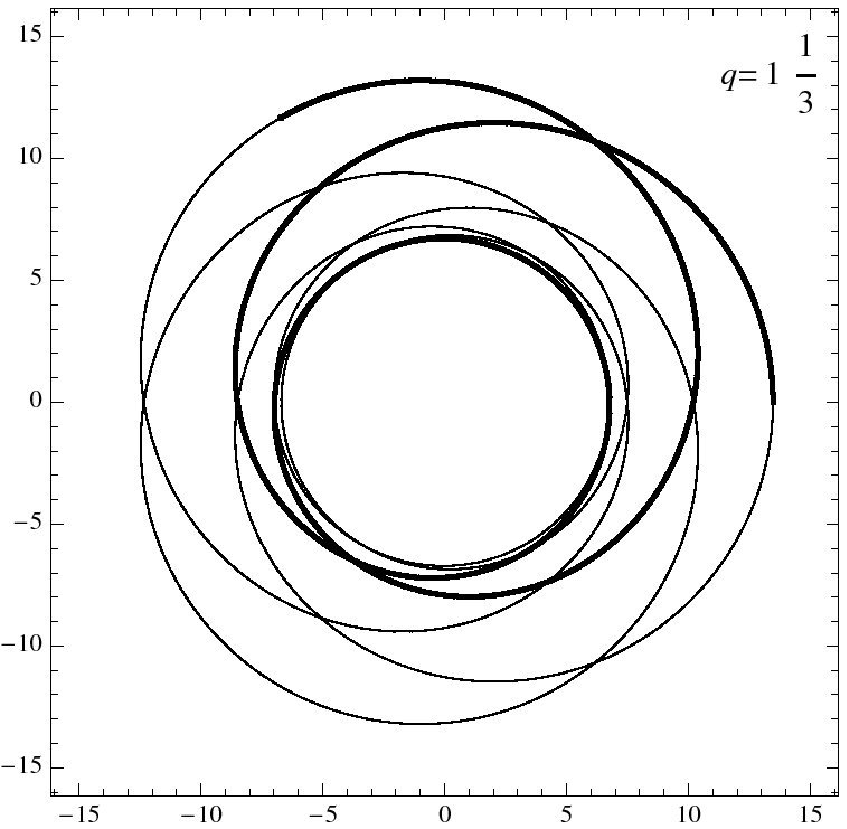}
\includegraphics[width=0.15\textwidth]{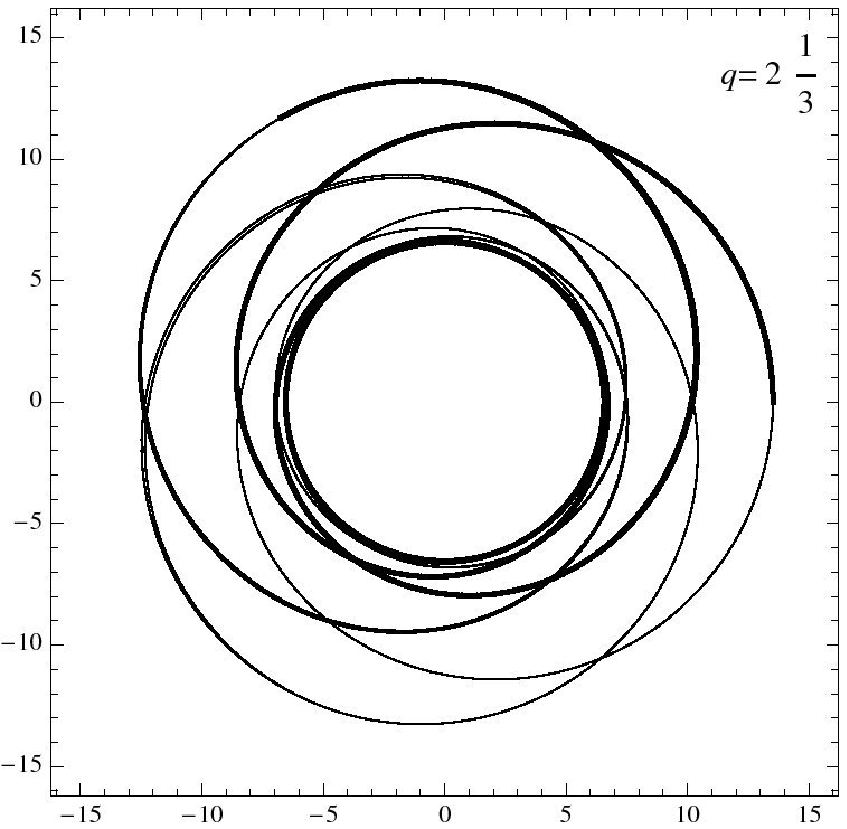}
\hfill


\includegraphics[width=0.15\textwidth]{xplots/000.eps}
\includegraphics[width=0.15\textwidth]{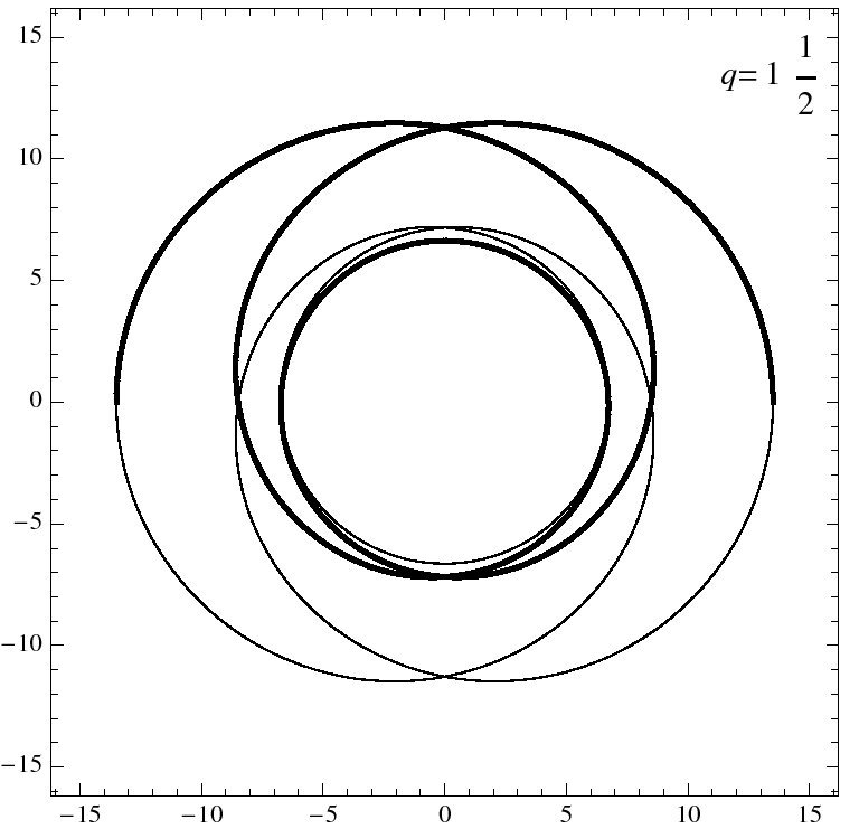}
\includegraphics[width=0.15\textwidth]{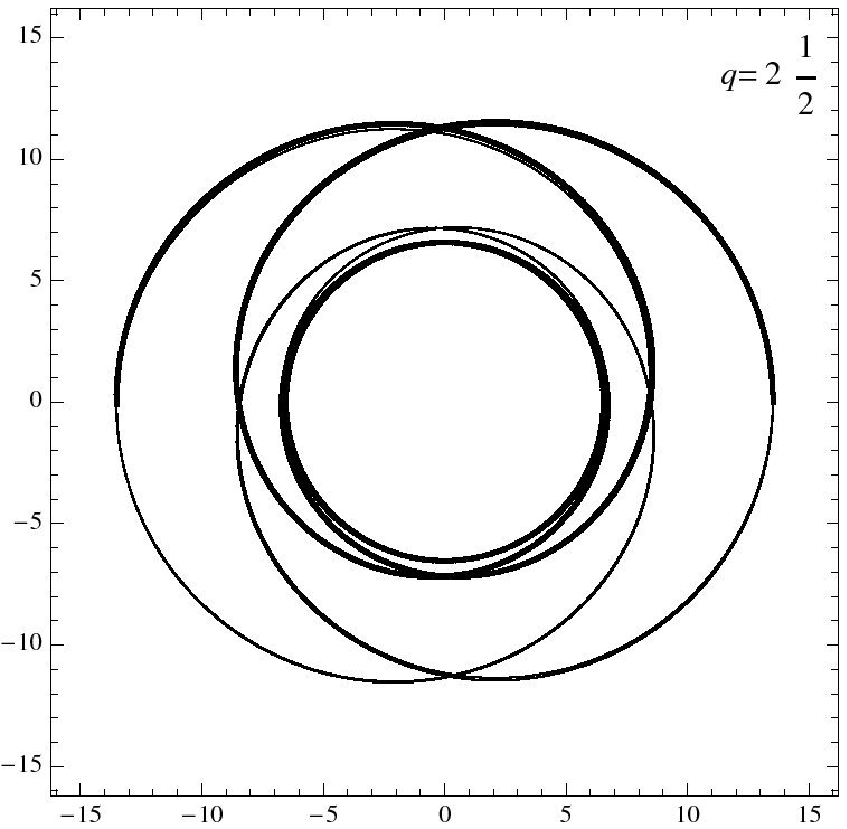}
\hfill


\includegraphics[width=0.15\textwidth]{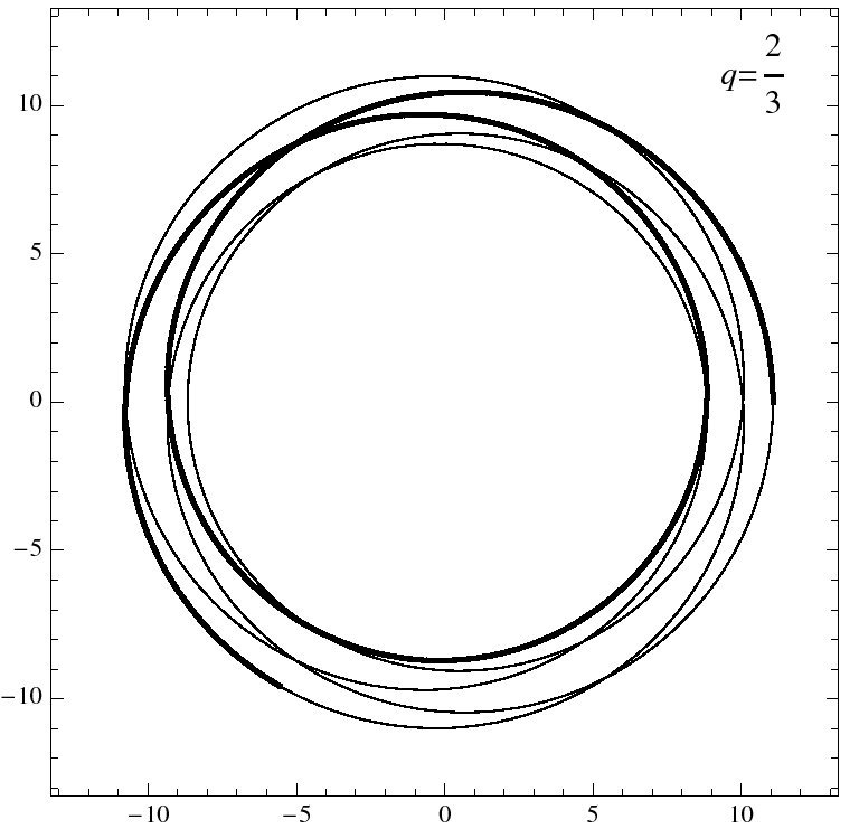}
\includegraphics[width=0.15\textwidth]{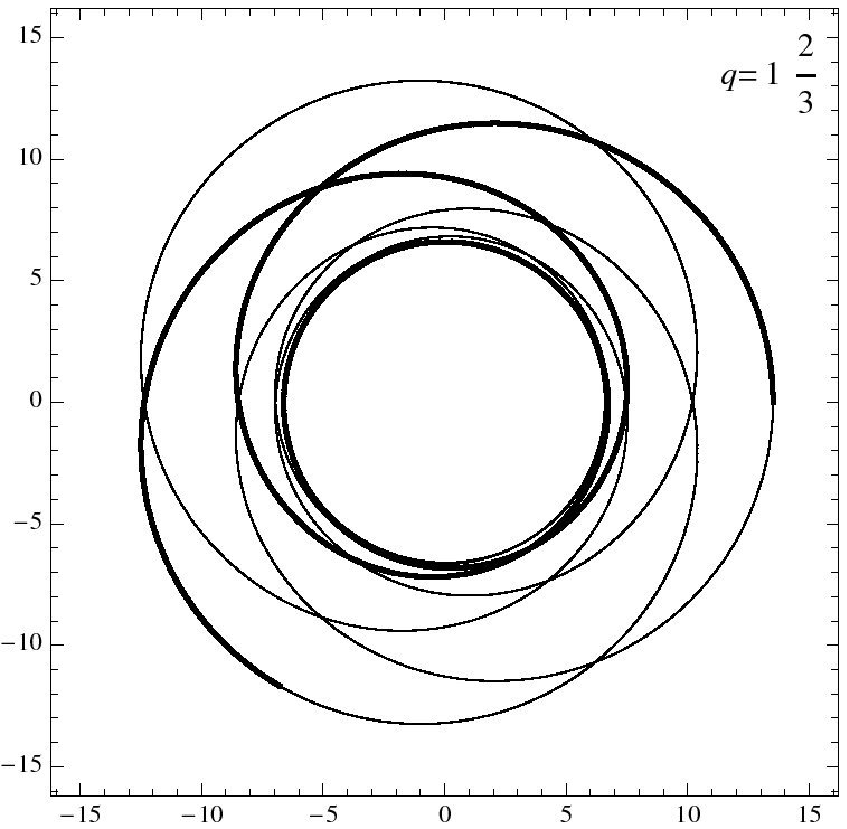}
\includegraphics[width=0.15\textwidth]{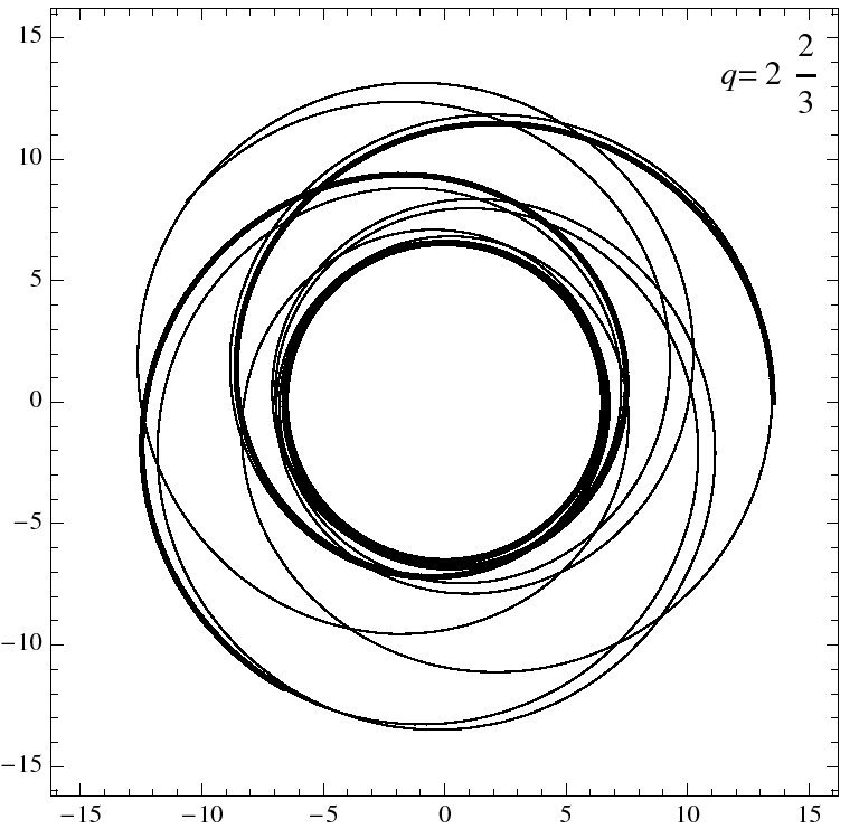}
\hfill
  \caption{A periodic table for orbits in a non-spinning system $a=0$,
  extreme-mass-ratio system,   $\frac{m_{2}}{m_{1}}=10^{-6}$. The
  angular momentum is
$L=3.9$ . Since $a=0$ all
  orbits lie in the equatorial plane. Periodic tables such as this one
  could be used to expand on comparisons with the full relativistic
  system.
All valid entries up to $z_\Phi=4$ are shown. As before, the final entry begins
  to show a departure from true periodicity as a result of numerical
  error. }
  \label{equatorial}
\end{figure}

The table of Fig.\ \ref{equatorial} reflects a non-spinning
black hole system with an extreme
mass ratio of $m_2/m_1=10^{-6}$. The first 3 entries are blank since $q_{min}$ is
just below $1/2$. Although we only show entries up to $z_\Phi=4$ for
$w_\Phi\le 2$, for these parameters $q_{max}=\infty$.
\ref{equatorial}. 
Although we defer a detailed
comparison between the 3PN spin-orbit system of this paper and the
Kerr system, we point out the intriguing possibility that periodic
tables could be used to further test the accuracy of the
Post-Newtonian expansion \cite{campanelli2008}.

\section{Conclusions}
\label{conc}

To recap in bullet format, for comparable mass
binaries with one spinning black hole and one non-spinning black hole
as approximated by the 3PN Hamiltonian plus spin-orbit
coupling, our main results are:

{\bf 1. Simplified Equations of Motion in an Orbital Basis}\par 
From which we find \par
$\bullet $ constant aphelia and perihelia for non-equatorial eccentric
orbits, and \par
$\bullet $ three fundamental frequencies that depend only on radius.\par

\noindent and

{\bf 2. Taxonomy of Fully Three-Dimensional Orbits}\par 
For which we find\par
$\bullet $ there exists a spectrum of closed orbits in the orbital
plane corresponding to a subset of the rationals;\par
$\bullet $ one rational, not two is required for an orbital plane
taxonomy of constant angular momentum slices;\par
$\bullet ${\it all} orbits can be approximated as near an
orbit that is
perfectly closed in the orbital plane; and\par
$\bullet $ zoom-whirl behavior is ubiquitous in comparable mass
binary dynamics and entirely quantifiable through the spectrum of rationals.

The first discovery we made in Ref.\ \cite{levin2008} with our periodic
taxonomy for equatorial Kerr orbits is that
precessing elliptical orbits such as Mercury's are excluded in the
strong-field Kerr regime -- just as Keplerian ellipses are excluded in any
relativistic regime.
Instead, at close separations, {\it all } equatorial 
Kerr eccentric orbits trace out precessions
of patterns best described as multi-leaf clovers, 
whirling around the central black hole
before zooming out quasi-elliptically \cite{{barack2004},{glampedakis2002}}. In
this paper we have
 found that this same conclusion applies in the strong-field regime
of our comparable mass black hole binaries, even out of the equatorial
plane. Our periodic tables in the
orbital plane show zoom-whirl behavior as the norm in the strong-field
regime and not as the exception.

The further importance of the orbital dynamics
lies in its direct imprint in
the gravitational waveform \cite{{levino2000},{drasco2006}}. The waveform will
necessarily reflect the features above. For instance, an equatorial
circular orbit (neglecting radiation reaction) is described
by essentially one frequency. By contrast, {\it all other orbits} in
the strong-field regime generate highly modulated waveforms 
naturally described by harmonics of the 3 orbital basis frequencies, which in turn
directly correspond to the natural
frequencies of a nearby periodic orbit.

Naturally, we should ask about the astrophysical likeliness of detecting
any such orbits with either the LIGO or LISA observatories. Although
estimates vary \cite{belczynski2007}, stellar mass black hole pairs are
currently the favored source for advanced ground-based dectectors and
intermediate mass black hole pairs are
considered an
important objective for space-based LISA science. It
is challenging to definitively assess the spins and eccentricities of
black hole/black hole binaries given the absence of observational
constraints \cite{oshaughnessy2005}.
Still, one can guess that long-lived stellar binaries that might
collapses to a pair of bound black holes would circularized by the
time the pair enters the strong-field due to
angular momentum lost in the form of gravitational
radiation.\footnote{However, when spin-spin coupling is included,
  there are no circular or
even spherical orbits \cite{companion}.}
By contrast, for shorter-lived black hole binaries formed in globular clusters,
the
astrophysical likeliness of eccentric orbits sliding in the LIGO
bandwidth
is
assessed to be $\gtrsim 30\%$ for eccentricities $>0.1$ in
Ref.\ \cite{wen2003}. All such binaries would necessarily transit near
the periodic set on inspiral. Even if the inspiral happens too
quickly to witness multiple executions near a low-leaf clover, the
orbit can still be sewn together
as a skip from a piece of one periodic to a piece of another.
Finally, the spins and eccentricities of intermediate black hole
binaries detectable by LISA are most
difficult to predict although we should expect them to spend a more
generous allotment of windings on eccentric orbits
in the strong-field. 

Before closing, we have to mention the effects of spin-spin coupling on
the orbital basis picture. The spin-spin correction introduces
additional precessions of the spin and this destroys the constancy of
the angle between $\bs\cdot \bl$ and generally introduces explicit
angular dependence in the equations of motion. Interpreted as a
small perturbation to the system here, the spin-spin couplings cause
additional wobbling of the precessional motion.  

When both objects spin, the impact of spin-spin coupling 
can be particularly destructive. It is by now well documented that two spinning
black holes in comparable mass binaries exhibit chaotic motion in the
conservative system
\cite{{suzuki1997},{levin2000},{levin2003},{cornish2002},{cornish2003},{hartl2005},{wu2007},{wu2008}}. 
There are not enough constants of the
motion to 
ensure regular behavior. 
Even the restricted spin-orbit scenario of this paper is 
clearly vulnerable to chaos even perturbed as it admits a homoclinic orbit. Under
perturbation, homoclinic orbits tend to disrupt into a homoclinic
tangle, an infinite intersection of stable and unstable
manifolds. At root, chaos emerges as
periodic orbits proliferate in a bounded region of space. Our
clean taxonomy of periodic orbits corresponding to a simple set
of rationals would give way to a glut of periodic orbits corresponding 
to some fractal set, as in systems with a strange repeller -- the
Hamiltonian analog to the strange attractor.
The complex motion may be
damped by energy and angular momentum loses to gravitational waves but,
at the least, chaos signals the breakdown of the simple set of
periodic orbits. The onset of chaos can be directly identified with
the spin-spin couplings, and we leave this task to a forthcoming
companion paper \cite{companion}.

\bigskip
\bigskip
\bigskip
\bigskip
\bigskip

\noindent{*Acknowledgments*}

We are especially thankful to Gabe Perez-Giz for his valuable and generous
contributions to this work and to Jamie Rollins for his careful
reading of the manuscript.
We also thank
Szabi Marka for important discussions.
JL and BG gratefully acknowledge the support of a Columbia University ISE grant.
This material is based in part upon work
supported under a National Science Foundation Graduate Research
Fellowship.

\vfill\eject
\appendix

\section{Projection of the equations of motion onto the non-orthogonal
  orbital basis}
\label{orbitalapp}

\subsection{Projection onto Orbital Basis}

By projecting the equations of motion onto the orbital basis
$(\bn,\bhPhi,\bhPsi)$, we show here that the equations of motion
depend only on the radius.

The four equations of motion in the orbital plane are obtained by
projecting Hamilton's equations onto the basis vectors, as is done in
celestial mechanics. For now,
consider only the projections onto the orbital basis vectors to
generate the four equations,
\begin{align}
\dot{\br}\cdot \bn&=\frac{\partial \H}{\partial {\bp}} \cdot \bn \nonumber \\
\dot{\br}\cdot \bhPhi&=\frac{\partial \H}{\partial {\bp}} \cdot \bhPhi\nonumber \\
\dot{\bp}\cdot \bn&= -\frac{\partial{\H}}{\partial {\br}}\cdot \bn \nonumber \\
\dot{\bp}\cdot \bhPhi&= -\frac{\partial{\H}}{\partial {\br}}\cdot \bhPhi
\quad .
\label{eomshort2pre}
\end{align}
To break down the LHS and RHS of the above projections it will be
useful to write
\begin{eqnarray}
\bp &=& (\bp\cdot \bn)\bn +(\bn\times \bp)\times \bn\\
 &=& P_r\bn +\frac{\bl}{r}\times \bn\\
 &=& P_r\bn +\frac{L}{r} \bhPhi
\label{pdecompose}
\end{eqnarray}
where the component 
$p_r=P_r$
and by capitol $P$'s we mean canonical momenta versus
small case, which will mean components.
To break down the LHS involves
\begin{align}
\dot{\br}&=\dot r \bn +r {\bf \dot{\bn}} \nonumber \\
\dot{\bp}&=\dot P_r \bn +P_r {\bf \dot{\bn}}-\frac{L}{r^2}\dot r
\bhPhi+\frac{L}{r}{\bf \dot{\bhPhi}} \quad .
\end{align}
So, we will need ${\bf \dot{\bn }}$ and ${\bf \dot{\bhPhi}}$, which are most
directly obtained by expanding $(\bn,\bhPhi)$ in the $(\bhx,\bhy)$
basis and then expanding $(\bhx,\bhy)$ in the Cartesian basis. So, we
will need
\begin{eqnarray}
\bn &=&\cos\Phi \bhx+\sin\Phi \bhy \nonumber \\
\bhPhi &=& -\sin\Phi \bhx +\cos \Phi \bhy \nonumber \\
\bhx &=& \cos\Psi \bhi +\sin\Psi \bhjy\nonumber \\
\bhy &=& \sin\theta_Y(
-\sin\Psi \bhi +\cos\Psi \bhjy)+\cos\theta_Y\bhk \nonumber \\
\bhPsi&=&-\sin\Psi \bhi +\cos\Psi \bhjy
\quad .
\label{need}
\end{eqnarray}
Using
\begin{eqnarray}
{\bf \dot{\bhx} }&=& \dot \Psi\bhPsi=\Omega_L \bhPsi \\
{\bf \dot{\bhPsi}} &=& -\dot\Psi\bhx =-\Omega_L \bhx \\
{\bf \dot{\bhy}} &=& \sin\theta_Y{\bf
  \dot{\bhPsi}}=-\sin\theta_Y\Omega_L \bhx 
\quad .
\end{eqnarray}
where $\dot\Psi=\Omega_L=\delta_1 J/r^3$ from Eq.\ (\ref{omegaL})
so that
we have ${\bf \dot {\bn}} $:
\begin{widetext}
\begin{eqnarray}
{\bf \dot{\bn}} &=&\dot \Phi\bhPhi+\cos\Phi {\bf \dot{\bhx}}+\sin\Phi
 {\bf \dot{\bhy }}\nonumber \\
 &=& 
\dot \Phi\bhPhi+\Omega_L\left (\cos\Phi {\bhPsi}-\sin\Phi\sin\theta_Y{\bhx }\right )\nonumber \\
&=&
\dot \Phi\bhPhi+\frac{\Omega_L}{\sin\theta_Y}\left (\cos\Phi
 {\bhy}-\cos\Phi\cos\theta_Y \bhk-\sin\Phi\sin^2\theta_Y{\bhx }\right
 )
\label{needdot}
\end{eqnarray}
\end{widetext}
To take the projection of Eqs.\ (\ref{eomshort2pre}) we will
also need
\begin{align}
{\bn} \cdot \bhx&= \cos\Phi \quad & \quad {\bhPhi} \cdot \bhx &= -\sin\Phi \nonumber \\
{\bn} \cdot \bhy&= \sin\Phi \quad &\quad {\bhPhi} \cdot \bhy &= \cos\Phi \nonumber \\
{\bn} \cdot \bhk&= \sin\Phi\cos\theta_Y \quad & \quad {\bhPhi} \cdot \bhk&= \cos\Phi\cos\theta_Y 
\label{nproject}
\end{align}
In the last step we use 
\begin{equation}
\bhk\cdot \bn=\bhj\cdot
(\cos\Phi\bhx+\sin\Phi\bhy)=\sin\Phi \bhj\cdot \bhy
\end{equation}
since $\bhx$ lies in the
equatorial plane, it is by definition perpendicular to $\bhj=\bhk$.
From all of the above relations we obtain for use in the projections
\begin{eqnarray}
{\bf \dot{\bn} }\cdot \bn&=& 0\\
{\bf \dot{\bn}} \cdot \bhPhi &=& \dot\Phi+\Omega_L\sin\theta_Y=
\dot\Phi+\Omega_L\cos\theta_L \quad .
\label{needproj}
\end{eqnarray}

Now for $\dot \bhPhi$. Taking the derivative of $\hat \Phi $ as
expressed in Eq.\ (\ref{need}) we have
\begin{widetext}
\begin{eqnarray}
{\bf \dot{\bhPhi}} &=& -
\dot \Phi\bn-\sin\Phi {\bf \dot{\bhx}}+\cos\Phi {\bf \dot{\bhy }}\nonumber \\
 &=& 
-\dot \Phi\bn+\Omega_L\left (-\sin\Phi {\bhPsi}-\cos\Phi\sin\theta_Y{\bhx }\right )\nonumber \\
&=&
-\dot \Phi\bn+\frac{\Omega_L}{\sin\theta_Y}\left (-\sin\Phi {\bhy}+\sin\Phi\cos\theta_Y \bhk-\cos\Phi\sin^2\theta_Y{\bhx }\right )
\end{eqnarray}
\end{widetext}
and using Eqs.\ (\ref{nproject}),
we have for use in the projections of Eqs.\ (\ref{eomshort2pre}), 
\begin{eqnarray}
{\bf \dot{\bhPhi}} \cdot \bn&=& -\left
(\dot\Phi+\Omega_L\sin\theta_Y\right )=
-\left (\dot\Phi+\Omega_L\cos\theta_L\right )
\nonumber \\
{\bf \dot{\bhPhi}} \cdot \bhPhi &=& 0\quad .
\end{eqnarray}

Now we can derive the equations of motion in the $(r,\Phi,\Psi)$
coordinates.
From the equations we constructed in section \S \ref{eomsection},
\begin{align}
\dot{\br}&=A\bp +B\bn +\delta_1 \frac{\bs\times \br}{r^3} \nonumber \\
\dot{\bp}&=C\bp +D\bn + \delta_1 \frac{\bs\times \bp}{r^3} 
+3\delta_1\frac{\bl\cdot \bs}{r^4} \bn
\quad ,
\label{eomshort2}
\end{align}
and
the projections (Eqs.\ (\ref{eomshort2pre})), with all of the
the above vector relations we have the radial equation,
\begin{equation}
\dot{\br}\cdot \bn=\dot r= AP_r +B \quad .
\end{equation}
The $\Phi$ equation is found from
\begin{eqnarray}
\dot{\br}\cdot \bhPhi &=&\frac{\partial H}{\partial \bp}\cdot \bhPhi \\
r\left (\dot \Phi +\Omega_L\cos\theta_L\right ) &=&
A \frac{L}{r}+
\delta_1 \frac{\left (\bs\times \br\right
  )\cdot \bhPhi}{r^3} \quad .
\label{rgamma}
\end{eqnarray}
Look at
\begin{eqnarray}
\left (\bs\times \br\right  )\cdot \bhPhi &=& 
r\left (\bs\cdot \bhl \right )\quad .
\end{eqnarray}
The $\Phi$ equation is then
\begin{eqnarray}
\dot \Phi
 &=& A\frac{L}{r^2}+\Omega_L \left
(-\cos\theta_L+\frac{\bs\cdot \bhl}{J}\right ) 
\label{group}
\end{eqnarray}
where $\bhs\cdot \bhl$ is constant.
Another helpful relation is 
\begin{eqnarray}
\frac{\bs\cdot\bhl}{J} &=& \bhj\cdot \bhl-\frac{L}{J}
=\cos\theta_L-\frac{L}{J}
\end{eqnarray}
allowing us to write the $\Phi$ equation in its final form,
\begin{eqnarray}
\dot \Phi
 &=& A\frac{L}{r^2} - \Omega_L \frac{L}{J} \quad .
\end{eqnarray}
The two conjugate momenta equations are next.
We start with $P_r$:
\begin{eqnarray}
\dot {\bp}\cdot \bn 
&=&\dot P_r
-\frac{L}{r}(\dot\Phi+\Omega_L\cos\theta_L) \\
&= & 
 C P_r +D 
+2\Omega_L
\frac{\bs\cdot \bl}{Jr} 
\nonumber
\end{eqnarray}
where we have used that
\begin{eqnarray}
(\bp \times \bs)\cdot \bn &=& \frac{\bs \cdot \bl}{r} 
\end{eqnarray}
Notice if we use Eq.\ (\ref{group}), we have 
\begin{eqnarray}
\dot P_r =
A\frac{L^2}{r^3}
+  
 C P_r +D +3\Omega_L\frac{\bs\cdot \bl}{Jr} 
\nonumber
\end{eqnarray}
and last
\begin{eqnarray}
\dot {\bp} \cdot \bhPhi &=& P_r(\dot\Phi+\Omega_L\cos\theta_L)
-\frac{L}{r^2}\dot r
\\
&=& C \frac{L }{r}
+\Omega_L
\frac{P_r \bs\cdot \bhl}{J}
\nonumber
\label{pgammaeom}
\end{eqnarray}
where we have used that
\begin{equation}
(\bp \times \bs)\cdot \bhPhi = \bs \cdot (\bhPhi
  \times \bp)=-P_r \bs\cdot \bhl
\end{equation}
Notice if we use Eq.\ (\ref{group}), we have a cancellation and
\begin{eqnarray}
 \left (A P_r -\dot r\right )\frac{L}{r^2} &=& -B\frac{L}{r}
= C \frac{L }{r}
\nonumber
\label{pgammaeom}
\end{eqnarray}
which confirms a true statement but does not provide any new equation of
motion since we implicitly used
$\dot L =0$. We will show in the next subection that the canonical momentum $P_\Phi=L$ 
and so the last equation of motion corresponds to $\dot P_\Phi=0$.
All four equations in the orbital basis are compiled in the boxed Eqs.\ (\ref{eoms}).

\subsection{Conjugate Momenta for $\Phi$ and $\Psi$}

We can show that the momentum conjugate to $\Phi$ is $P_\Phi=L$ and
the momentum conjugate to $\Psi$ is $P_\Psi=L_z=L\cos\theta_L$. 
So the equations of motion $(\Phi,P_\Phi)$ and $(\Psi,P_\Psi)$ should be
derivable from 
\begin{align}
\dot\Phi=\frac{\partial \H}{\partial P_\Phi}\quad , \quad \quad \dot
P_\Phi&=0\nonumber \\
\dot\Psi=\frac{\partial \H}{\partial P_\Psi}\quad , \quad \quad \dot
P_\Psi&=0
\label{hamconj}
\end{align}
This is far more elaborate than one might guess and so we spend this
last subsection verifing that $L$ and $L_z$ are the conjugate momenta
and that the equations of motion derived according to (\ref{hamconj})
are in fact the same as those of Eqs.\ (\ref{eoms}).

We begin by showing that $P_\Phi=L$ and $P_\Psi=L_z$ are consistent
with our equations before we explicitly rederive the equations of motion
using (\ref{hamconj}). Bear
in mind that 
the variables $(r,\Phi,\Psi)$ and their conjugate momenta must be
linearly independent and so $\partial X^i/\partial X^j=\delta^i_j$
where $X=(r,P_r,\Phi,P_\Phi,\Psi,P_\Psi)$. 
We also need to be careful to rewrite everything in terms of
$(r,\Phi,\Psi)$ and the conjugate momenta $(P_r,L,L_z)$. Particularly,
we will need to take the derivative of terms like $\cos\theta_L=P_\Psi/P_\Phi=L_z/L$.
From
Eqs.\ (\ref{hamconj}) we have that
\begin{align}
\dot\Phi &=\frac{\partial \H}{\partial P_\Phi}=\frac{\partial
  \H}{\partial \bp}\cdot
\frac{\partial \bp}{\partial P_\Phi}+
\frac{\partial
  \H}{\partial \br}\cdot
\frac{\partial \br}{\partial P_\Phi}\nonumber \\
&=\dot\br\cdot
\frac{\partial \bp}{\partial L}-\dot\bp
\frac{\partial \br}{\partial L}
\label{stepup1}
\end{align}
Now
\begin{align}
\dot\br&=\dot r \bn+r{\bf\dot{\bn}}\nonumber \\
\dot\bp&=\dot P_r \bn+P_r{\bf\dot{\bn}}-\frac{L\dot
  r}{r^2}\bhPhi+\frac{L}{r}{\bf \dot{\bhPhi}}
\quad .
\label{stepup3}
\end{align}
and
\begin{align}
\frac{\partial \bp}{\partial L}&=P_r
\frac{\partial \bn}{\partial L}+\frac{L}{r}\frac{\partial
  \bhPhi}{\partial L}+\frac{1}{r}\bhPhi \nonumber \\ 
\frac{\partial \br}{\partial L}&=r\frac{\partial \bn}{\partial L}
\quad .
\label{stepup2}
\end{align}
The unit vectors $\bn$ and $\bhPhi$ depend on $L$ and $L_z$ through
$\cos\theta_L=L_z/L$.
Using Eq.\ (\ref{need}) and replacing $\sin\theta_Y$ with
$\cos\theta_L$ etc. we have
\begin{align}
\frac{\partial \bn}{\partial L}&=\sin\Phi\frac{\partial \bhy}{\partial
  L}=-
\sin\Phi\frac{\cos\theta_L}{L}\left ( \bhPsi-\cot\theta_L\bhk\right
  )\nonumber \\
\frac{\partial \bhPhi}{\partial L}&=\cos\Phi\frac{\partial \bhy}{\partial
  L}=-
\cos\Phi\frac{\cos\theta_L}{L}\left ( \bhPsi-\cot\theta_L\bhk\right
  )
\end{align}
Using Eq.\ (\ref{needdot}) for ${\bf\dot{\bn}}$ and Eq.\
(\ref{needproj}) for ${\bf\dot{\bn}}\cdot\bhPhi$,
Eq.\ (\ref{stepup1}) becomes
\begin{widetext}
\begin{align}
\dot\Phi=
\dot\Phi+\dot\Psi\cos\theta_L+
\frac{\cos\theta_L}{L} 
\left (\cot\theta_L\bhk-\bhPsi\right )\cdot
\left [\left (P_r\sin\Phi+\frac{L}{r}\cos\Phi \right)\left (\dot r\bn+r{\bf \dot\bn}\right )
-
r\sin\Phi\left (\dot P_r\bn+P_r{\bf \dot\bn}-\frac{L\dot r}{r^2}
\bhPhi+
\frac{L}{r}{\bf \dot\bhPhi}\right )\right ]\nonumber
\end{align}
\end{widetext}
Taking the dot products, there are some fortuante cancellations and overall we find
\begin{equation}
\dot\Phi=\dot\Phi+\dot\Psi\cos\theta_L-\dot\Psi\cos\theta_L=\dot\Phi
\quad .
\end{equation}
As claimed, $P_\Phi=L$ is consistent. 
The same procedure for $\Psi$ with $P_\Psi=L_z$ yields a similarly
consistent equality.

Now, to derive the equations of motion directly from 
\begin{equation}
\dot\Phi=\frac{\partial\H}{\partial L}=
\frac{\partial\H_{PN}}{\partial L}+
\frac{\partial\H_{SO}}{\partial L}
\end{equation}
is tedious but doable. The PN piece is straightforward if one first
writes everything in terms of canonical variables; i.e.,
$(\bn\cdot\bp)=P_r$ and $\bp^2=P_r^2+P_\Phi^2/r^2=P_r^2+L^2/r^2$, then
\begin{equation}
 \frac{\partial\H_{PN}}{\partial L}=A\frac{L}{r^2}\quad \quad .
\label{HpnPhi}
\end{equation}
The
partial of the spin orbit piece is
\begin{align}
\frac{\partial\H_{SO}}{\partial L}&=
\frac{\delta_1 \bs}{r^3}\cdot \frac{\partial (\br\times \bp)}{\partial
  L}  \nonumber \\
&=
\frac{\delta_1 \bs}{r^3}\cdot \left [r\frac{\partial\bn} {\partial L}\times\bp+
\br\times \frac{\partial\bp }{\partial L} \right ]\nonumber \\
&=
\frac{\delta_1 \bs}{r^3}\cdot \left [r\frac{\partial\bn }{\partial L}\times\bp+
\br\times \left (P_r\frac{\partial\bn }{\partial L} +\frac{L}{r}
\frac{\partial\bhPhi }{\partial L} +\frac{\bhPhi}{r}\right )\right ]
\nonumber
\end{align}
Replace $P_r\bn$ with $\bp-(L/r)\bhPhi$ to cancel the first
cross product with part of the second cross product. We then have
\begin{align}
\frac{\partial\H_{SO}}{\partial L}
=\delta_1\frac{ \bs\cdot \bhl}{r^3}+
\frac{\delta_1L }{r^3}\bs\cdot \left [
\left (\bn\times\frac{\partial\bhPhi }{\partial L} \right)
-
\left(\bhPhi\times \frac{\partial\bn }{\partial L}\right )
\right ]
\nonumber
\end{align}
Taking the cross products we have
\begin{align}
\frac{\partial\H_{SO}}{\partial L}
&=\delta_1\frac{ \bs\cdot \bhl}{r^3}-
\frac{\delta_1 }{r^3}\bs\cdot
\cos\theta_L\left (\bhk-\cot\theta_L\bhPsi\right )
\nonumber\\
&=\delta_1\frac{ \bs\cdot \bhl}{r^3}-
\frac{\delta_1 }{r^3}
\cos\theta_L\left (J-L_z-\cot\theta_L\bs\cdot \bhPsi\right )
\nonumber
\end{align}
where in the last step we have used $\bs\cdot \bhk=J-L_z$.
Taking $\bs\cdot
\bhPsi=(\bj-\bl)\cdot\bhPsi=-\bl\cdot\bhPsi=-L\sin\theta_L$
gives
\begin{align}
\frac{\partial\H_{SO}}{\partial L}
&=\delta_1\frac{ \bs\cdot \bhl}{r^3}-
\frac{\delta_1 }{r^3}
\cos\theta_L\left (J-L_z-L\cos\theta_L\right )
\nonumber \\
&=\delta_1\frac{ \bs\cdot \bhl}{r^3}-
\frac{\delta_1 J}{r^3}
\cos\theta_L
\nonumber \\
&=-
\Omega_L\frac{L}{J}
\quad \quad .
\label{HSOPhi}
\end{align}
Added together Eqs.\ (\ref{HpnPhi}) and (\ref{HSOPhi}) give the
equation of motion for $\Phi$ in Eqs.\ (\ref{eoms}) as claimed.
The $\dot\Psi$ equation can be derived similarly. Since
both $P_\Phi$ and $P_\Psi$ are constants, the Hamiltonian is cyclic in
$\Phi$ and $\Psi$.

\bibliographystyle{aip.bst}
\bibliography{pntax}

\end{document}